\documentclass[twocolumn]{aastex62}
\hypersetup{linkcolor=red,citecolor=green,filecolor=cyan,urlcolor=magenta}

\def\ts     {\thinspace} 
\usepackage{amsmath}

\def\kms  {\ifmmode{{\rm \ts km\ts s}^{-1}}\else{\ts km\ts s$^{-1}$\ts}\fi}
\def\msol {\ifmmode{{\rm M}_{\odot}}\else{M$_{\odot}$\ts}\fi}
\def\lsun {\ifmmode{{\rm L}_{\odot}}\else{L$_{\odot}$\ts}\fi}
\def\cii  {\ifmmode{{\rm [C}{\rm \scriptstyle II}]}\else{[C\ts {\scriptsize II}]\ts}\fi}
\def\ci   {\ifmmode{{\rm C}{\rm \scriptstyle I}}\else{C\ts {\scriptsize I}\ts}\fi}
\def\m    {\ifmmode{\mu {\rm m}}\else{$\mu$m}\fi}
\def\hi   {\ifmmode{{\rm H}{\rm \scriptstyle I}}\else{H\ts {\scriptsize I}\ts}\fi}
\def\hii  {\ifmmode{{\rm H}{\rm \scriptstyle II}}\else{H\ts {\scriptsize II}\ts}\fi}
\def\nii  {\ifmmode{{\rm [N}{\rm \scriptstyle II}]}\else{[N\ts {\scriptsize II}]\ts}\fi}
\def\oiii {\ifmmode{{\rm [O}{\rm \scriptstyle III}]}\else{[O\ts {\scriptsize III}]\ts}\fi}
\def\hh   {\ifmmode{{\rm H}_2}\else{H$_2$\ts}\fi}
\def\nhh  {\ifmmode{N({\rm H}_2)}\else{$N$(H$_2$)\ts}\fi}
\def\microns {\ifmmode{\mu{\rm m}}\else{$\mu$m\ts}\fi}

\graphicspath{{./}{figures/}}

\shorttitle{The ALMA Spectroscopic Survey in the HUDF: Deep 1.2 mm continuum number counts}
\shortauthors{Gonz\'alez-L\'opez et al.}

\begin{document}

\title{The ALMA Spectroscopic Survey in the HUDF: Deep 1.2 mm continuum number counts}

\correspondingauthor{Jorge Gonz\'alez-L\'opez}
\email{jgonzalez@carnegiescience.edu}

\author[0000-0003-3926-1411]{Jorge Gonz\'alez-L\'opez}
\affil{Las Campanas Observatory, Carnegie Institution of Washington, Casilla 601, La Serena, Chile}
\affil{N\'ucleo de Astronom\'ia de la Facultad de Ingenier\'ia y Ciencias, Universidad Diego Portales, Av. Ej\'ercito Libertador 441, Santiago, Chile}
\nocollaboration

\author[0000-0001-8695-825X]{Mladen Novak}
\affil{Max Planck Institute f\"ur Astronomie, K\"onigstuhl 17, 69117 Heidelberg, Germany}

\author[0000-0002-2662-8803]{Roberto Decarli}
\affil{INAF-Osservatorio di Astrofisica e Scienza dello Spazio, via Gobetti 93/3, I-40129, Bologna, Italy}

\author[0000-0003-4793-7880]{Fabian Walter}
\affil{Max Planck Institute f\"ur Astronomie, K\"onigstuhl 17, 69117 Heidelberg, Germany}
\affil{National Radio Astronomy Observatory, Pete V. Domenici Array Science Center, P.O. Box O, Socorro, NM 87801, USA}

\author[0000-0002-6290-3198]{Manuel Aravena}
\affil{N\'ucleo de Astronom\'ia de la Facultad de Ingenier\'ia y Ciencias, Universidad Diego Portales, Av. Ej\'ercito Libertador 441, Santiago, Chile}

\author{Chris Carilli}
\affil{National Radio Astronomy Observatory, Pete V. Domenici Array Science Center, P.O. Box O, Socorro, NM 87801, USA}
\affil{Battcock Centre for Experimental Astrophysics, Cavendish Laboratory,
Cambridge CB3 0HE, UK}

\author[0000-0002-3952-8588]{Leindert Boogaard}
\affil{Leiden Observatory, Leiden University, PO Box 9513, NL-2300 RA Leiden, The Netherlands}

\author{Gerg\"{o} Popping}
\affil{Max Planck Institute f\"ur Astronomie, K\"onigstuhl 17, 69117 Heidelberg, Germany}
\affil{European Southern Observatory, Karl-Schwarzschild-Strasse 2, 85748, Garching, Germany}

\author{Axel Weiss}
\affil{Max-Planck-Institut f\"ur Radioastronomie, Auf dem H\"ugel 69, 53121 Bonn, Germany}

\author{Roberto J. Assef}
\affil{N\'ucleo de Astronom\'ia de la Facultad de Ingenier\'ia y Ciencias, Universidad Diego Portales, Av. Ej\'ercito Libertador 441, Santiago, Chile}

\author[0000-0002-8686-8737]{Franz Erik Bauer}
\affil{Instituto de Astrof\'{\i}sica, Facultad de F\'{\i}sica, Pontificia Universidad Cat\'olica de Chile Av. Vicu\~na Mackenna 4860, 782-0436 Macul, Santiago, Chile}
\affil{Millennium Institute of Astrophysics (MAS), Nuncio Monse{\~{n}}or S{\'{o}}tero Sanz 100, Providencia, Santiago, Chile}
\affil{Space Science Institute, 4750 Walnut Street, Suite 205, Boulder, CO 80301, USA}

\author{Rychard Bouwens}
\affil{Leiden Observatory, Leiden University, PO Box 9513, NL-2300 RA Leiden, The Netherlands}

\author{Paulo C.~Cortes}
\affil{Joint ALMA Observatory - ESO, Av. Alonso de C\'ordova, 3104, Santiago, Chile}
\affil{National Radio Astronomy Observatory, 520 Edgemont Rd, Charlottesville, VA, 22903, USA} 

\author{Pierre Cox}
\affil{Institut d'astrophysique de Paris, Sorbonne Universit\'e, CNRS, UMR 7095, 98 bis bd Arago, 7014 Paris, France}

\author{Emanuele Daddi}
\affil{Laboratoire AIM, CEA/DSM-CNRS-Universite Paris Diderot, Irfu/Service d'Astrophysique, CEA Saclay, Orme des Merisiers, 91191 Gif-sur-Yvette cedex, France}

\author{Elisabete da Cunha}
\affil{International Centre for Radio Astronomy Research, University of Western Australia, 35 Stirling Hwy, Crawley, WA 6009, Australia}
\affil{Research School of Astronomy and Astrophysics, Australian National University, Canberra, ACT 2611, Australia}
\affil{ARC Centre of Excellence for All Sky Astrophysics in 3 Dimensions (ASTRO 3D)}

\author{Tanio D\'iaz-Santos}
\affil{N\'ucleo de Astronom\'ia de la Facultad de Ingenier\'ia y Ciencias, Universidad Diego Portales, Av. Ej\'ercito Libertador 441, Santiago, Chile}

\author{Rob Ivison}
\affil{European Southern Observatory, Karl-Schwarzschild-Strasse 2, 85748, Garching, Germany}
\affil{Institute for Astronomy, University of Edinburgh, Royal Observatory, Blackford Hill, Edinburgh EH9 3HJ}

\author{Benjamin Magnelli}
\affil{Argelander-Institut f\"ur Astronomie, Universit\"at Bonn, Auf dem H\"ugel 71, 53121 Bonn, Germany}

\author[0000-0001-9585-1462]{Dominik Riechers}
\affil{Cornell University, 220 Space Sciences Building, Ithaca, NY 14853, USA}
\affil{Max Planck Institute f\"ur Astronomie, K\"onigstuhl 17, 69117 Heidelberg, Germany}
\affil{Humboldt Research Fellow}

\author{Ian Smail}
\affil{Centre for Extragalactic Astronomy, Department of Physics, Durham University, South Road, Durham, DH1 3LE, UK}

\author{Paul van der Werf}
\affil{Leiden Observatory, Leiden University, PO Box 9513, NL-2300 RA Leiden, The Netherlands}

\author{Jeff Wagg}
\affil{SKA Organization, Lower Withington Macclesfield, Cheshire SK11 9DL, UK}

\begin{abstract}

We present the results from the 1.2 mm continuum image obtained as part of the ALMA Spectroscopic Survey in the Hubble Ultra Deep Field (ASPECS). The 1.2 mm continuum image has a size of 2.9 (4.2) arcmin$^2$ within a primary beam response of 50\% (10\%) and a rms value of $9.3\ts{\rm\mu Jy\ts beam^{-1}}$. We detect 35 {sources} at high significance (Fidelity $\geq0.5$), 32 of these have well characterized near-infrared HST counterparts.

We estimate the 1.2 mm number counts to flux levels of $<30\ts{\rm\mu Jy}$ in two different ways: we first use the detected sources to constrain the number counts and find a significant flattening of the counts below $S_\nu \sim 0.1$ mJy. In a second approach, we constrain the number counts by using a probability of deflection statistics (P(D)) analysis. For this latter approach, we describe new methods to accurately measure the noise in interferometric imaging (employing jack-knifing in the cube and in the visibility plane). This independent measurement confirms the flattening of the number counts. 
Our analysis of the differential number counts shows that we are detecting $\sim$93\% ($\sim$100\% if we include the lower fidelity detections) of the total continuum dust emission associated to galaxies in the HUDF.

The ancillary data allows us to study the dependence of the 1.2 mm number counts on redshift ($z=0\text{--}4$), galaxy dust mass (${\rm M}_{\rm dust}=10^{7}\text{--}10^{9}\msol$), stellar mass (${\rm M}_{*}=10^{9}\text{--}10^{12}\msol$), and star-formation rate (${\rm SFR}=1\text{--}1000\ts\msol\ts{\rm yr^{-1}}$). In an accompanying paper we show that the number counts are crucial to constrain galaxy evolution models and the understanding of star-forming galaxies at high redshift. 

\end{abstract}



\section{Introduction} \label{sec:Introduction}

In order to explain the number of stars and galaxies we see in the local Universe, a large population of star-forming galaxies must have been present in the past \citep{Madau_Dickinson2014}. The stellar radiation produced by such young galaxies will be partially absorbed by interstellar dust and re-emitted in the mid and far infrared (FIR). The combination of the emission of all galaxies at different times then produces a cosmic infrared background (CIB). 

The Cosmic Background Explorer (COBE) satellite detected the CIB in multiple wavelengths and concluded that the observed emission should correspond to dust reprocessed emission from high-z galaxies \citep{Puget1996,Hauser1998,Fixsen1998,HauserDwek2001}. 

Soon after the detection of the CIB, bolometer camera observations in submm bands revealed a population of dust enshrouded highly star-forming galaxies at high redshift \citep{Smail1997,Barger1998,Hughes1998,Eales1999}. These galaxies are bright at submm-wavelengths and barely detected in the UV/Optical bands, hence their name submillimeter galaxies or dusty star-forming galaxies (DSFGs). 
The discovery of this population of galaxies showed that a considerable fraction of the star-formation activity in high redshift galaxies could be obscured by dust \citep[e.g. an early review by][]{Blain2002}. 

As expected, Atacama Large Millimeter/Submillimeter Array (ALMA) is revolutionizing the study of DSFGs. First by allowing for very high angular resolution observations of bright DSFGs \citep[e.g.][]{ALMA2015,Iono2016,Hodge2016,Tadaki2018} and also for allowing the detection of the faint high-redshift population \citep{Watson2015,Laporte2017,Hashimoto2019}. It is not a surprise that some of the observations made with ALMA were to follow-up single-dish submm sources \citep[e.g.][]{Karim2013,Hodge2013,Simpson2015,Stach2018}. 
The next step was to do deep ALMA surveys to search for the population of faint DSFGs. 
These first attempts focused on fields with deep archival data such as Subaru/XMM-Newton Deep Survey Field \citep[SXDF,][]{Hatsukade2013,Hatsukade2016}, SSA22 \citep{Umehata2017}, GOODS-S/{\it Hubble} Ultra
Deep Field \citep[HUDF,][]{Walter2016,Aravena2016,Dunlop2017,Franco2018,Hatsukade2018}, Frontier Fields \citep{GL2017,MunozArancibia2018}, on calibrator fields \citep{Oteo2016} and on combined multiple single-pointing fields \citep{Ono2014,Carniani2015,Oteo2016,Fujimoto2016}.

The flux density distribution of DSFGs is a powerful tool to test galaxy evolution models. Straight-forward measurements, such as the galaxies number counts, are the result of several intrinsically complex processes. In order to model the observed number counts of galaxies at submm and mm wavelengths we need to take into account the dark matter halo distribution \citep[e.g.][]{Klypin2011,Klypin2016}, the star-formation history and modes \citep[e.g.][]{Sargent2012,Bethermin2012}, spectral energy distribution (SED) of galaxies \citep[e.g.][]{daCunha2013a}, redshift distribution \citep[e.g.][]{Bethermin2015}, type of observations \citep[e.g.][]{MunozArancibia2015,Bethermin2017} and the observed distribution of galaxies at all wavelengths \citep[e.g.][]{Schreiber2017}. 
In order to test the galaxy evolution models, we need to obtain reliable number counts of well characterized DSFGs.

In this work we present the band 6 (1.2 mm) continuum observations obtained as part of the ALMA Spectroscopic Survey in the HUDF \citep[ASPECS,][]{Walter2016}. The HUDF, and specially the eXtreme Deep Field (XDF), is the deepest extragalactic fields observed by the {\it Hubble} space telescope \citep{Beckwith2006,Illingworth2013}.  
{ASPECS Large Program (ASPECS-LP)} is an ALMA cycle 4 large program that represents an unparalleled three-dimensional survey in a contiguous $\sim4$ arcmin$^2$ region in the HUDF designed to trace the cosmic evolution of cool gas and dust.
The first results using the band 3 observations were recently published \citep{GL2019,Decarli2019,Boogaard2019,Aravena2019a,Popping2019}. 
By collapsing the band 6 data we obtain a deep continuum image that can be used to search for the faint population of DSFGs (hereafter referred to as 1mm galaxies). We present the number counts of sources detected in the deep continuum image of the HUDF and how they change for different galaxy properties.  

This paper is structured as follows; in \S\ref{sec:Observations} we present the observations, the calibration and imaging process. In \S\ref{sec:Methods} we describe the methods used to extract the sources as well as to estimate the number counts. In \S\ref{sec:Results} we present the results from the source extraction and in \S\ref{sec:Discussion} we discuss the implications of such results. Finally, in
\S\ref{sec:Conclusion} we present the our conclusions. 

{Throughout this paper, the properties of the galaxies were estimated adopting a \citet{Chabrier2003} initial mass function and a flat $\Lambda$CDM cosmology as in \citet{Boogaard2019,Aravena2019a,Aravena2019b}.}

\section{Observations and data processing} \label{sec:Observations}

\begin{figure*}
\epsscale{1.2}
\plotone{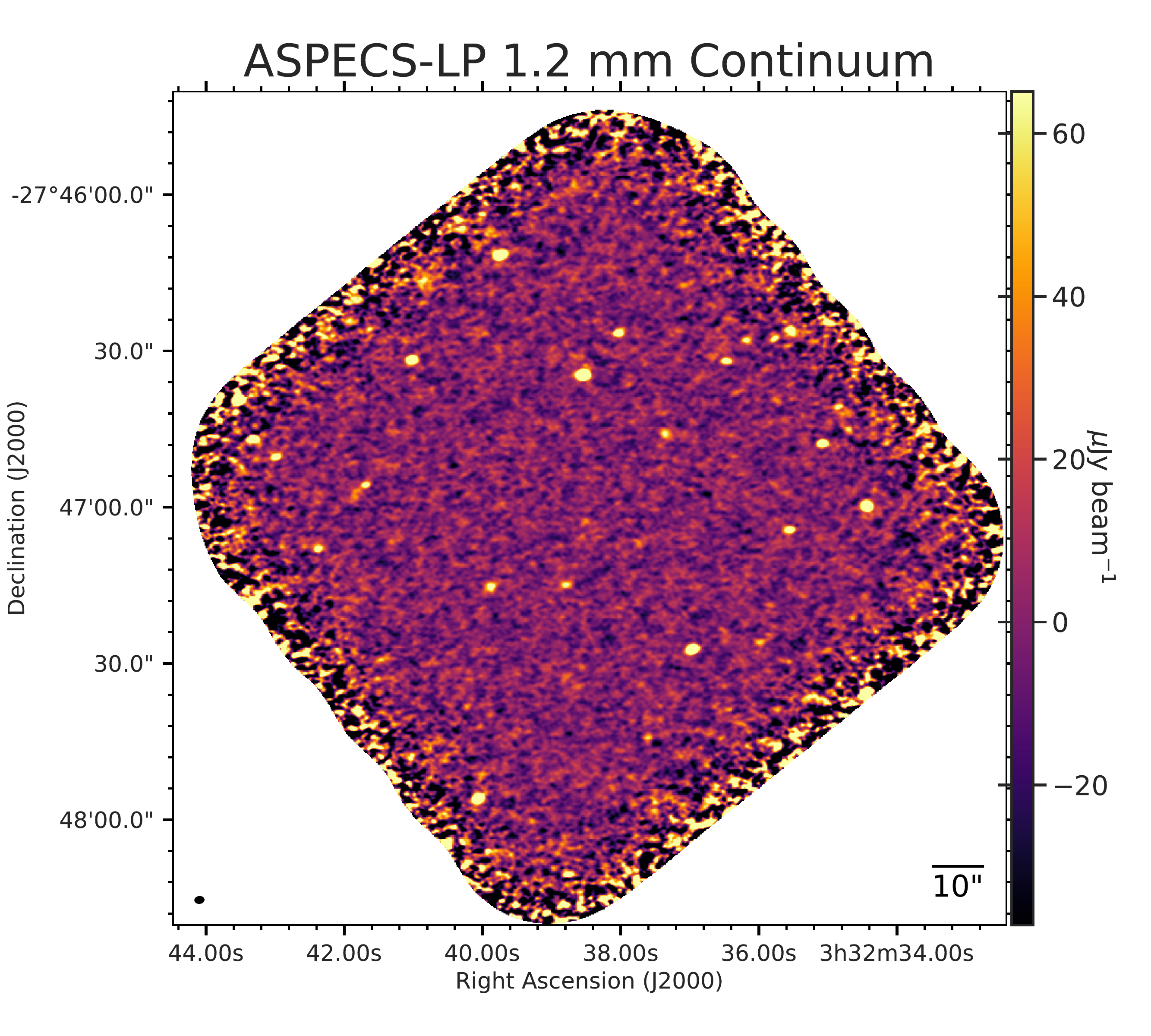}
\caption{1.2 mm continuum image in the H--UDF obtained as part of ASPECS-LP. The image is obtained using natural weighting and it is corrected by the mosaic primary beam response. The total area shown here corresponds to 4.2 arcmin$^2$ (PB$\geq0.1$). The synthesized beam is shown in bottom left corner.\label{fig:ContinuumMap}}
\end{figure*}

\subsection{Survey design}

The data used in this work correspond to the band 6 observations from ASPECS-LP. The observational setups used in ASPECS-LP are the same as the ones used in ASPECS-Pilot observations presented in \citet{Walter2016}. {We used eight spectral tunings that cover most of the ALMA band 6 (212--272 GHz, $\approx94\%$ of band 6). The mosaic consists of 85 pointings separated by $11\arcsec$ ($\approx51\%$ of the half power beam width at the highest frequency setup) and is Nyquist-sampled at all frequencies. }
The ASPECS-LP observations were designed to obtain a sensitivity similar to that of the ASPECS-Pilot observations. The final data ended up being deeper than requested owing to several of the executions being obtained with excellent weather conditions and low precipitable water vapor (${\rm pwv}<1{\rm \ts mm}$). 

\subsection{Data reduction, calibration and imaging}

\begin{figure*}
\epsscale{1.2}
\plotone{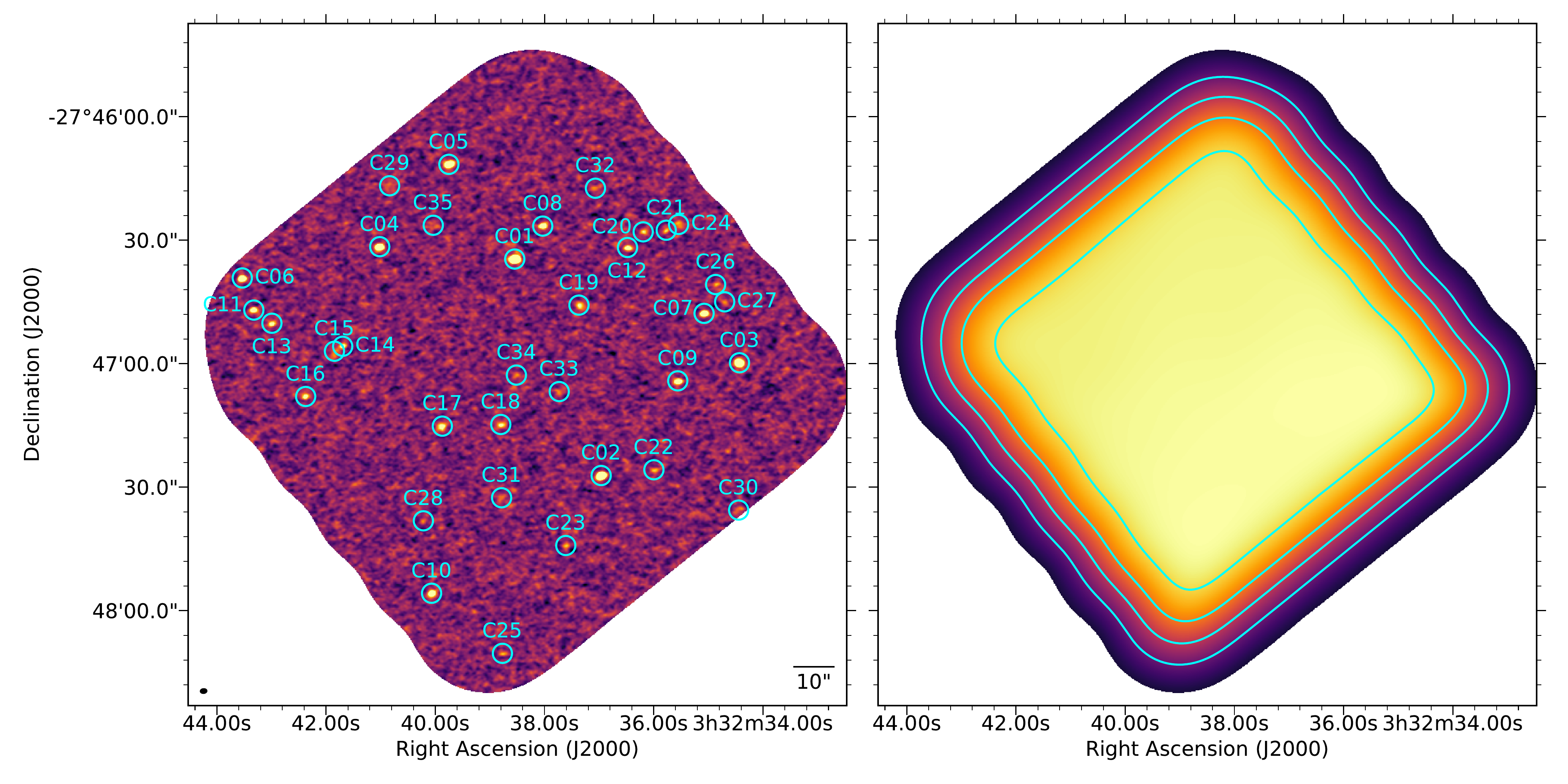}
\caption{The left panel shows the continuum image without mosaic primary beam correction. Cyan circles show the positions and IDs of the source candidates found to be significant. The right panel shows the primary beam response of the continuum image mosaic. The contours show where the primary beam response is 0.3, 0.5, 0.7 and 0.9, respectively. The total image has a size of 4.2 arcmin$^2$, while the area within the PB$\geq$ 0.3, 0.5, 0.7 and 0.9 are 3.4, 2.9, 2.4 and 1.8 arcmin$^2$ respectively. \label{fig:ContinuumMapSources}}
\end{figure*}

We processed the data with the {\sc CASA} ALMA calibration pipeline \citep[v.5.1.1;][]{McMullin2007}, using the calibration and flagging input provided by ALMA. 
We created the continuum images as well as the data cubes with {\sc CASA} v.5.4.0-70. That version of {\sc CASA} fixed a couple of bugs that affected large mosaic imaging. We obtained the continuum images using the task {\sc tclean} with natural weighting and multi-frequency synthesis (mfs) mode. We created the cleaned images by putting cleaning boxes on all the sources with $\rm{S/N}\geq5$ and cleaning down to $20\ts{\rm \mu Jy\ts beam^{-1}}$. 
The deepest region of the continuum image has a rms value of $9.3\ts{\rm \mu Jy\ts beam^{-1}}$ with a beam size of $1.53\arcsec\times1.08\arcsec$. We used the same procedure to obtain a tapered image with a rms value of $11.3\ts{\rm \mu Jy\ts beam^{-1}}$ with a beam size of $2.37\arcsec\times2.05\arcsec$. We used the tapered image to search for extended emission that could be missed by the original natural weighting image. {We chose this tapered beam size because it offers extra sensitivity to slightly extended emission without a significant loss of depth. The resulting beam is similar to the band 3 continuum image beam of $\approx2.1\arcsec$ \citep{GL2019}.}

In addition to the continuum image, we created the data cube covering the frequency range. We did identify those spaxels of each continuum source that contain bright emission lines. We discovered several emission lines associated with the continuum detected sources, with some sources showing up to 3 emission lines in band 6. Because of this, we proceeded to create a line-free continuum image by flagging all the frequency channels with strong emission lines \citep{GL2019,Decarli2019}. We repeated the same procedure used for the original cleaned continuum image, resulting in a rms of $10.0\ts{\rm \mu Jy\ts beam^{-1}}$. This image will be used to look for flux density boosting produced by the emission lines. 

\section{Methods} \label{sec:Methods}

\subsection{Source search}\label{sec:SourceSearch}

We performed 2D source extraction using LineSeeker in the same manner as was done for the ASPECS-LP 3 mm continuum image \citep{GL2019}.
{LineSeeker uses a simple algorithm to extract sources either in collapsed channel maps created for the search of emission lines or in continuum images. In both cases the extraction is based on the source peak flux density per beam. Similar extraction algorithms have been used in previous studies \citep{Aravena2016,GL2017}.}
{The source extraction was done in the natural weighting and tapered images independently. } We estimated an initial rms value from the image, we excluded from the image all the pixels with $\rm{S/N}\geq5$ and estimated a new rms value. We selected all the pixels with new $\rm{S/N}\geq2$ and grouped them together using Density-Based Spatial Clustering of Applications with Noise (DBSCAN) algorithm \citep{Ester1996}. DBSCAN is useful to recover the emission of extended sources as long as such emission is traced by pixels with $\rm{S/N}\geq2$. At the same time, two blended sources will be associated to the same source if they are connected by pixels with $\rm{S/N}\geq2$. Visual inspection is needed to determine if blending is occurring for any source. 
LineSeeker automatically estimated a fidelity value for each source in the output catalog by comparing the number of positive and negatives with similar signal-to-noise ratios as follows

\begin{equation}
{\rm Fidelity}=1 - \frac{N_{\rm Neg}}{N_{\rm Pos}},
\end{equation}

with $N_{\rm Neg}$ and $N_{\rm Pos}$ being the number of negative and positive continuum candidates detected with a given S/N value. 

{We estimated the completeness correction by injecting point sources of different flux density values into $uv$-plane jackknife noise reference image (described in \S\ref{sec:NoiseReferenceImage}). In each iteration we injected 20 point sources convolved by the corresponding clean synthesized beam. We chose to inject only 20 sources at a time to decrease the probability of blending. The process was repeated until we injected 20,000 sources. We then used LineSeeker to extract the sources from the new images. We classified each of the injected sources as recovered if they are detected with a signal-to-noise value higher than a given limit associated with fidelity equal to $50\%$ ($\rm S/N=4.3$ for the natural image and $\rm S/N=3.3$ for the tapered image).}
{We used the same images to estimate the flux--boosting effect on the recovered sources. We estimate an excess of $\approx11\%$ in the measured flux density for the sources detected with fidelity equal to $50\%$. In both images the flux density excess decreased to $\lesssim3\%$ at $\rm S/N\sim5$.}
{The injection of the sources was done in the image-plane instead of the $uv$-plane. Accordingly to \href{https://library.nrao.edu/public/memos/naasc/NAASC_117.pdf}{NAASC Memo \#117} the flux recovery of sources in ALMA mosaic images should be reliable. }

{The final list of sources detected with fidelity equal or higher than $50\%$ in the natural weighted or the tapered images is presented in Tab. \ref{tab:LPContinuum}. For the sources detected in both images the fidelity value was selected as the highest of the two. In order to account for possible extended faint galaxies, all completeness correction values were taken from the tapered image analysis.}
{As reference, we detect 27 positive sources with $\rm S/N\geq4.3$ and 2 negative sources with the same $\rm S/N$ in the natural weighted image. In the case of the  tapered image we detect 32 positive sources with $\rm S/N\geq3.3$ and 3 negative sources.  }

\subsection{Direct number counts}
\label{sec:DirectNumberCounts}

The most common method for estimating the number counts is to directly count the detected sources and make corrections for the fidelity and completeness. We followed the same recipe used by \citet{Aravena2016} and \citet{GL2019}, where the number counts per bin ($N(S_{i})$) are computed as follows

\begin{equation}
    N(S_{i}) = \frac{1}{A}\sum_{j=1}^{X_{j}}\frac{P_{j}}{C_{j}},
\end{equation}

where $A$ is the total area of the observations ($1.16\times10^{-3}\ts{\rm deg^{2}}$ for ${\rm PB}\geq0.1$), $P_{j}$ is the probability of each source being real (Fidelity) and $C_{j}$ is the completeness correction for the corresponding intrinsic flux density. Note that we estimated the completeness correction on the mosaic primary beam corrected plane, so the information about the different sensitivity across the map is included in this correction factor and not in an associated effective area per source.
To account for extended emission being missed by the natural weighting image, we used the correction factor estimates from the tapered image for all sources. These correction factors are $\approx1.3\text{--}1.4\times$ larger in the faint end (30--40 $\mu$Jy) and practically equal for sources brighter than 0.1 mJy. 
{The flux density values for all the sources are corrected by flux--boosting effects (described in \S\ref{sec:SourceSearch}) at the moment of estimating the number counts. }

We obtained the cumulative number counts by summing up each $N(S_{i})$ over all the possible $\geq{\rm  S}_{i}$. The size of the bins is $\log{S_{\nu}}=0.25$ for the differential number counts and $\log{S_{\nu}}=0.1$ for the cumulative number counts. We used all source candidates with ${\rm Fidelity}\geq0.5$ listed in Tab. \ref{tab:LPContinuum} for the number counts. 
{The error estimates for the number counts were done by combining Poisson statistics errors based on the number of sources per bin and the intrinsic uncertainty of the flux density measurements. For the latter we generated new flux density values for each source following a Gaussian distribution given by the corresponding flux density estimates and their errors. We then measured the number counts using these new flux density values. We repeated this process 1000 times and measured the scatter per bin. For simplicity, this scatter is added in quadrature to the Poisson statistics errors. }
{We checked that the main results from the direct number counts do not change if we apply a more restrictive cut in PB. We repeated the analysis using ${\rm PB}\geq0.5$ and ${\rm PB}\geq0.9$ (the latter corresponding to approximately half of the total area) and the number counts in the faint end remained the same. The only difference appears in the bright end where the error bars increase because of the smaller area used.}

\subsection{Number counts}\label{sec:P_D_Analysis}

\begin{figure}
\epsscale{1.2}
\plotone{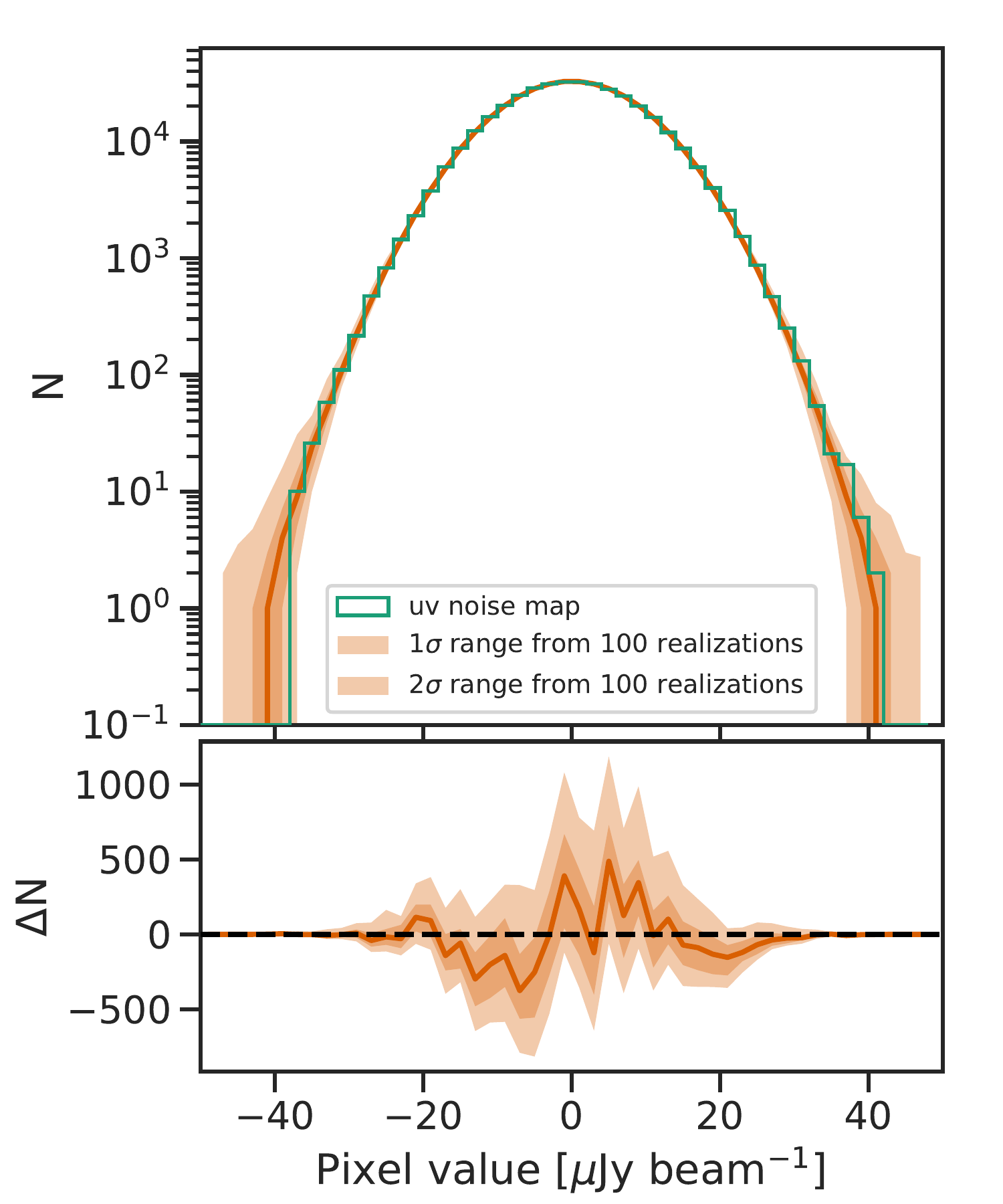}
\caption{Top panel: Comparison of the pixel distribution of the noise image created from the $uv$-plane jackknife and the dirty cube jackknife (for details see \S\ref{sec:P_D_Analysis}). Bottom panel: Residual between the $uv$-plane jackknife and the dirty cube jackknife. \label{fig:NoiseMapsMFS}}
\end{figure}

\begin{figure}
\epsscale{1.2}
\plotone{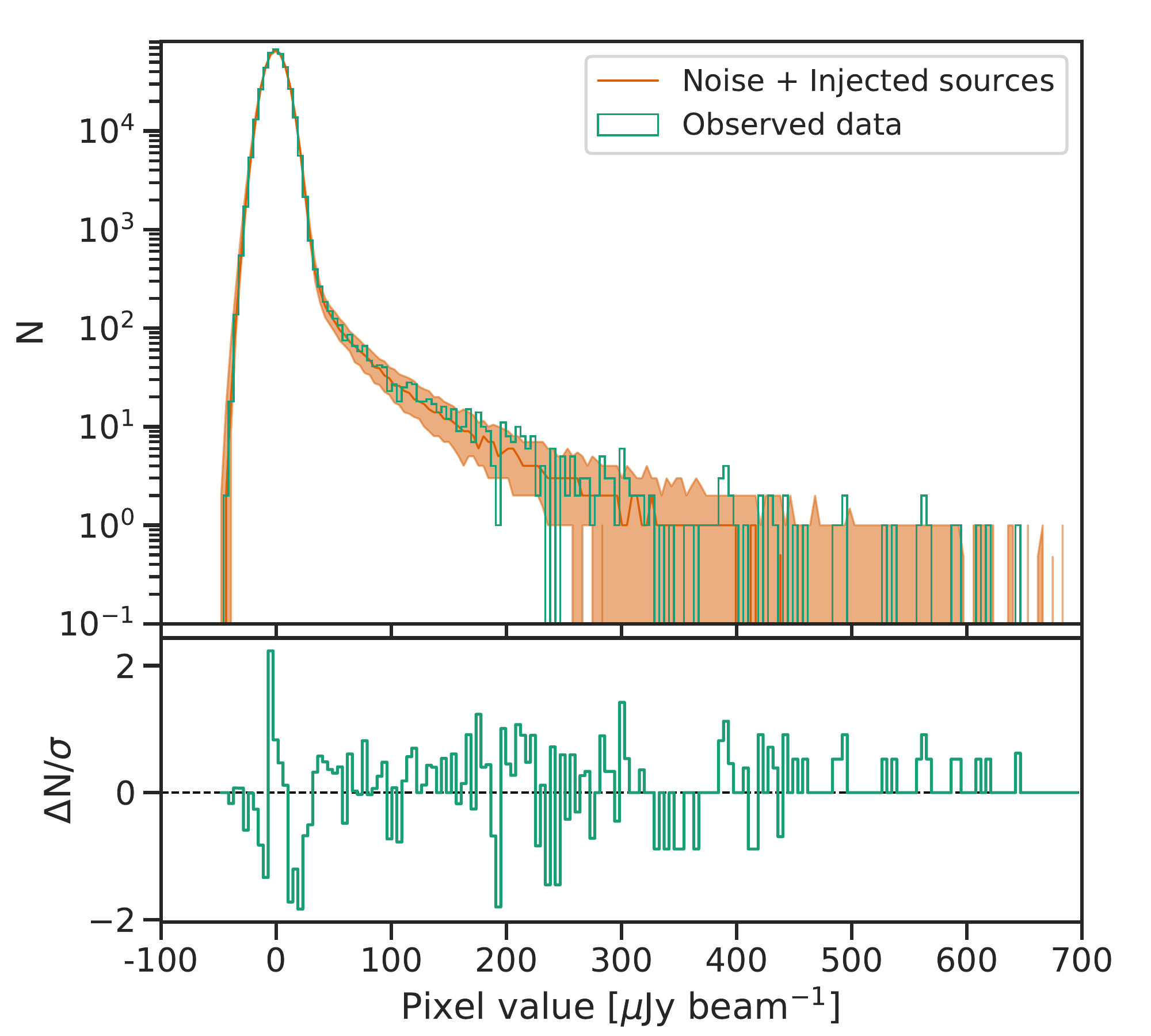}
\caption{Top panel: Green histogram shows the distribution of pixels in the natural images without primary beam correction. The orange line and region show the median distribution of pixels obtained from combining the $uv$-plane noise and injected sources following the best fit distribution (see details in \S\ref{sec:P_D_Analysis}). Bottom panel: Residual between the data and the best fit shown in red in the top panel. 
\label{fig:PixelDistribution}}
\end{figure}

\subsubsection{Input needed for analysis}

An alternative method to estimate the intrinsic distribution of sources in the observed data is to compare the observed image pixel distribution to what would be produced by a given assumed number counts distribution. This is called probability of deflection statistics or P(D) analysis \citep{Condon1974} and has been used to estimate the number counts from noise confused bolometer camera observations \citep[e.g. ][]{Hughes1998,Weiss2009}. 

To do this P(D) analysis, or forward modeling, we needed to know the dirty beam, the mosaic primary beam response and the intrinsic noise properties of the observations. We produced the first two maps with {\sc tclean} at the imaging stage. The one missing piece was the intrinsic noise properties of the observations. 
We could not use the residual map created by {\sc tclean} as noise reference because it can be contaminated by sources not bright enough to be selected for cleaning boxes and by confusion noise \citep{Condon1974,Scheuer1974}. We needed a method to remove all the real signal emission from the observations. 

\subsubsection{Noise reference images}
\label{sec:NoiseReferenceImage}

We here present two independent approaches to obtain a noise reference image from a calibrated ALMA dataset. The first is based on the jackknife resampling used in single-dish bolometer observations \citep{Perera2008,Scott2008,Weiss2009}. If the emission is stable with time, then the image produced by the combination of the jackknifed scans (where the amplitudes in every second scan are multiplied by $-1$) will have all the real emission removed, whereas the noise properties are conserved. This will provide a good representation of the noise properties of the observations. 
The same exact jackknife resampling can not be used with interferometric observations since different scans do not provide information from the same region on the sky. Each interferometric integration corresponds to one point in the $uv$-plane and is sensitive to one determined position and angular scale on the sky. To bypass this limitation we jackknifed the observations by inverting (multiply the real and imaginary part by $-1$ or shift phase by $180\degr$) every second visibility, which artificially creates destructive interference on the real sky emission.

The second approach to obtain a noise reference image is also based on jackknife resampling but done in channel space of the data cube. 
In a sense, all continuum images made from ALMA observations are done by combining the information from multiple channels into one. Such combination can be done by fitting a continuum emission to all different channels, as done by {\sc tclean} in mfs mode, or by collapsing the associated data cube. 
Based on this, we randomly inverted (multiply by $-1$) half of the channels in the dirty cube created for the band 6 ASPECS-LP data. We repeated this process 100 times, in order to obtain several realization of the noise. For each jackknifed dirty cube, we created a continuum noise image by taking the weighted average of all channels. 
We calculated the weights by channels as $w=1/\sigma^2_i$, with $\sigma_i$ being the rms of each individual channel. 
This weight scheme should be similar to the continuum image produced by {\sc tclean} in mfs mode, which uses the weights calculated during calibration and re-normalized during the concatenation of the different executions. 

In Fig. \ref{fig:NoiseMapsMFS} we present the histograms of the $uv$-plane noise reference and the one and two sigma distributions from the channels jackknife resampling. Both independent jackknife processes return very similar distributions. The rms measured in the $uv$-plane noise is $9.21\ts{\rm \mu Jy\ts beam^{-1}}$ while the median and one sigma range for rms values from the channel jackknifing images is $9.03_{-0.03}^{+0.04}\ts{\rm \mu Jy\ts beam^{-1}}$. Based on these results, we can be confident that the real noise distribution of our data is in the range $9.0-9.2\ts{\rm \mu Jy\ts beam^{-1}}$. 
As reference, when using weights of $w=1/\sigma_i$ for the weighted average on the continuum image creation, the resulting rms values from the channel jackknife is $9.16\pm0.06\ts{\rm \mu Jy\ts beam^{-1}}$, in agreement with the rms value obtained from $uv$-plane noise. 
We conclude that the differences produced by the different weighting schemes are in the sub-$\mu$Jy range. 

With the dirty beam, the mosaic primary beam response and the intrinsic noise properties of the observations, we proceeded to conduct the forward modeling of the number counts from the observed data. 

\subsubsection{P(D) analysis}

For the fitting process we assumed that the differential number counts can be well described by a double power-law function \citep{Scott2002,Franco2018}, as given by

\begin{equation}
\label{eq:DPL}
    \frac{dN}{dS} = \frac{N_{0}}{S_{0}}\left[\left(\frac{S}{S_{0}}\right)^{\alpha}+\left(\frac{S}{S_{0}}\right)^{\beta} \right]^{-1},
\end{equation}

with the explored ranges for the different parameters being $N_{0}=1\text{--}100\times10^{3}{\rm \ts mJy^{-1}deg^{-2}}$, $S_{0}=0.05\text{--}1.0{\rm \ts mJy}$, $\alpha=1.5\text{--}5$ and $\beta$ between $-2$ and $+2$. {Previous studies have used this double power-law function to model the number counts break seen at $\sim4\text{--}5$ mJy at $870\mu m$ and 1.1 mm \citep{Franco2018,Stach2018}. Given the area and depth of our observations we expect to detected sources with flux densities values $<1$ mJy. In this regime the number counts have been observed to follow a single power-law \citep{Fujimoto2016,Franco2018}. The chosen parameter space is designed to test if the single power-law function is a good description of the number counts at the faint end probed in our observations or if another break is needed.} 

For a given set of parameters {(sampled from a grid of parameters)} for the differential number counts and the area of the image, we created a list of flux density values between 0.03 and 2 mJy following such distribution. We injected point sources with the corresponding flux density value into a blank image at random positions. We then corrected the flux density for each source by the mosaic primary beam response at the corresponding position. Once all the sources were injected, the blank image was convolved with the dirty beam and added to the noise reference image.

This process was repeated 1000 times to obtain the pixel distribution of noise plus injected sources for a given combination of parameters. 
We compared the different iterations distributions with the pixel histogram obtained of the real dirty image with bins of $4.3\ts{\rm \mu Jy\ts beam^{-1}}$ and calculated the likelihood of that model being a good representation of the data using

\begin{equation}
    L = \prod_{i=1}^{Nbins}\sum_{m=0}^{M} m^{N_i}\frac{\mathrm{e}^{-m}}{N_i!}P_i(m),
\end{equation}

with $Nbins$ being the number of bins in the histogram, $N_i$ is the number of pixels in each bin for the real dirty image, $m$ is the expected number of pixels in each bin for a given combination of parameters and $P_i(m)$ is the probability density distribution of getting the value $m$ in the iterations. 
The likelihood is the product of the Poisson probability of measuring the observed number of pixels per bin given an expected rate $m$, which follows the distribution from the 1000 iterations. 

For simplicity, instead of maximizing L, we minimize the associated quantity given by

\begin{equation}
    C = -2\ln{L}, 
\end{equation}

which is the maximum likelihood-based statistic for Poisson data with a prior for the expected rate \citep[][]{Cash1979}. 

In Fig. \ref{fig:PixelDistribution} we show the observed pixel distribution (from the dirty image) compared to the best-fit from the forward modeling number counts. In the bottom panel we show the residuals of the subtraction of the best model from the observed histogram. 
The $\sigma$ value used in the residuals was obtained by combining in quadrature the Poisson statistical errors associated to the observed number of pixels within a bin and the central $68\%$ range of the distribution of values from the 1000 repetition of the injection of sources ($P(m)$). The scatter associated to the 1000 repetitions dominates with respect to the Poisson statistical error. This is because a given source falling in a region with different mosaic primary beam values will result in a different pixel distribution. 

The fitting of the number counts was done with the dirty image instead of the cleaned image. By doing so we included the effects of the dirty beam negative sidelobes on the negative pixels distribution. The complete distribution of pixels is broadened by the negative sidelobes of the dirty beam and the confusion noise \citep{Weiss2009}.

\section{Results} \label{sec:Results}


\begin{deluxetable*}{cccccccc}
\tablecaption{Continuum source candidates in ASPECS-LP 1.2 mm continuum image. \label{tab:LPContinuum}}
\tablehead{
\colhead{ID} & 
\colhead{R.A.} & 
\colhead{Dec} & 
\colhead{S/N} & 
\colhead{Fidelity} & 
\colhead{PBC} & 
\colhead{$S_{1.2\,\,mm}$} & 
\colhead{HST?} \\
\colhead{ASPECS-LP.1mm.} & 
\colhead{} & 
\colhead{} & 
\colhead{} & 
\colhead{} & 
\colhead{} & 
\colhead{[$\rm \mu Jy$]} & 
\colhead{}
}
\colnumbers
\startdata
C01 & 03:32:38.54 & $-$27:46:34.60 & 67.6 & $1.0_{-0.0}^{+0.0}$ & 0.97 & $752.0 \pm 24.2$ & Yes \\
C02 & 03:32:36.96 & $-$27:47:27.20 & 44.1 & $1.0_{-0.0}^{+0.0}$ & 0.99 & $431.6 \pm 9.8$ & Yes \\
C03 & 03:32:34.43 & $-$27:46:59.79 & 30.7 & $1.0_{-0.0}^{+0.0}$ & 0.86 & $429.2 \pm 23.0$ & Yes \\
C04 & 03:32:41.02 & $-$27:46:31.60 & 26.8 & $1.0_{-0.0}^{+0.0}$ & 0.82 & $316.2 \pm 11.8$ & Yes \\
C05 & 03:32:39.75 & $-$27:46:11.60 & 23.2 & $1.0_{-0.0}^{+0.0}$ & 0.66 & $461.2 \pm 28.2$ & Yes \\
C06 & 03:32:43.53 & $-$27:46:39.19 & 22.6 & $1.0_{-0.0}^{+0.0}$ & 0.2 & $1071.0 \pm 47.4$ & Yes \\
C07 & 03:32:35.08 & $-$27:46:47.80 & 20.1 & $1.0_{-0.0}^{+0.0}$ & 0.84 & $232.6 \pm 11.5$ & Yes \\
C08 & 03:32:38.03 & $-$27:46:26.60 & 16.2 & $1.0_{-0.0}^{+0.0}$ & 0.96 & $163.2 \pm 10.1$ & Yes \\
C09 & 03:32:35.56 & $-$27:47:04.20 & 15.9 & $1.0_{-0.0}^{+0.0}$ & 1.0 & $154.6 \pm 9.7$ & Yes \\
C10 & 03:32:40.07 & $-$27:47:55.80 & 13.8 & $1.0_{-0.0}^{+0.0}$ & 0.49 & $342.1 \pm 33.8$ & Yes \\
C11 & 03:32:43.32 & $-$27:46:46.99 & 13.6 & $1.0_{-0.0}^{+0.0}$ & 0.46 & $288.8 \pm 21.2$ & Yes \\
C12 & 03:32:36.48 & $-$27:46:31.80 & 10.7 & $1.0_{-0.0}^{+0.0}$ & 0.91 & $113.7 \pm 10.6$ & Yes \\
C13 & 03:32:42.99 & $-$27:46:50.19 & 9.7 & $1.0_{-0.0}^{+0.0}$ & 0.66 & $116.0 \pm 15.6$ & Yes \\
C14a & 03:32:41.69 & $-$27:46:55.80 & 9.4 & $1.0_{-0.0}^{+0.0}$ & 0.95 & $95.9 \pm 9.8$ & Yes \\
C14b & 03:32:41.85 & $-$27:46:57.00 & 9.4 & $1.0_{-0.0}^{+0.0}$ & 0.95 & $89.2 \pm 19.7$ & Yes \\
C15 & 03:32:42.37 & $-$27:47:08.00 & 8.9 & $1.0_{-0.0}^{+0.0}$ & 0.73 & $118.0 \pm 13.2$ & Yes \\
C16 & 03:32:39.87 & $-$27:47:15.20 & 8.8 & $1.0_{-0.0}^{+0.0}$ & 0.98 & $142.8 \pm 17.6$ & Yes \\
C17 & 03:32:38.80 & $-$27:47:14.80 & 8.1 & $1.0_{-0.0}^{+0.0}$ & 0.99 & $96.9 \pm 15.3$ & Yes \\
C18 & 03:32:37.37 & $-$27:46:45.80 & 7.2 & $1.0_{-0.0}^{+0.0}$ & 0.98 & $107.2 \pm 16.1$ & Yes \\
C19 & 03:32:36.19 & $-$27:46:28.00 & 6.8 & $1.0_{-0.0}^{+0.0}$ & 0.78 & $84.6 \pm 12.4$ & Yes \\
C20 & 03:32:35.77 & $-$27:46:27.60 & 6.0 & $1.0_{-0.0}^{+0.0}$ & 0.62 & $94.5 \pm 15.7$ & Yes \\
C21 & 03:32:36.00 & $-$27:47:25.80 & 5.5 & $1.0_{-0.0}^{+0.0}$ & 0.92 & $58.3 \pm 10.5$ & Yes \\
C22 & 03:32:37.61 & $-$27:47:44.20 & 5.5 & $1.0_{-0.0}^{+0.0}$ & 0.91 & $58.8 \pm 10.7$ & Yes \\
C23 & 03:32:35.55 & $-$27:46:26.20 & 5.4 & $1.0_{-0.0}^{+0.0}$ & 0.48 & $147.5 \pm 29.8$ & Yes \\
C24 & 03:32:38.77 & $-$27:48:10.40 & 5.4 & $1.0_{-0.0}^{+0.0}$ & 0.39 & $134.5 \pm 24.9$ & Yes \\
C25 & 03:32:34.87 & $-$27:46:40.80 & 5.4 & $1.0_{-0.0}^{+0.0}$ & 0.58 & $90.0 \pm 16.8$ & Yes \\
C26 & 03:32:34.70 & $-$27:46:45.00 & 4.3 & $0.54_{-0.17}^{+0.17}$ & 0.64 & $65.3 \pm 15.2$ & Yes \\
C27 & 03:32:40.22 & $-$27:47:38.20 & 4.1 & $0.78_{-0.08}^{+0.08}$ & 0.85 & $46.4 \pm 11.3$ & No \\
C28 & 03:32:40.84 & $-$27:46:16.80 & 3.9 & $0.87_{-0.03}^{+0.02}$ & 0.44 & $184.1 \pm 45.8$ & Yes \\
C29 & 03:32:34.45 & $-$27:47:35.60 & 3.5 & $0.8_{-0.04}^{+0.04}$ & 0.13 & $307.8 \pm 75.3$ & No \\
C30 & 03:32:38.79 & $-$27:47:32.60 & 3.5 & $0.8_{-0.04}^{+0.04}$ & 1.0 & $34.1 \pm 9.7$ & Yes \\
C31 & 03:32:37.07 & $-$27:46:17.40 & 3.5 & $0.8_{-0.04}^{+0.04}$ & 0.84 & $47.4 \pm 11.5$ & Yes \\
C32 & 03:32:37.73 & $-$27:47:06.80 & 3.5 & $0.8_{-0.04}^{+0.04}$ & 0.99 & $40.5 \pm 9.8$ & Yes \\
C33 & 03:32:38.51 & $-$27:47:02.80 & 3.3 & $0.55_{-0.09}^{+0.1}$ & 0.98 & $41.8 \pm 9.8$ & Yes \\
C34 & 03:32:40.04 & $-$27:46:26.40 & 3.3 & $0.55_{-0.09}^{+0.1}$ & 0.91 & $38.7 \pm 10.7$ & No \\
\enddata
\tablecomments{
(1) Identification for continuum source candidates discovered in ASPECS-LP 1.2 mm continuum image.
(2) Right ascension (J2000).
(3) Declination (J2000).
(4) S/N value obtained by LineSeeker assuming an unresolved source.
(5) Fidelity estimate using negative detection and Poisson statistics.
(6) Mosaic primary beam response.
(7) Integrated flux density at 1.2  mm obtained after removing the channels with bright emission lines when necessary.
(8) Presence of HST counterpart. Details in \citet{Aravena2019b}.
}
\end{deluxetable*}

Table \ref{tab:LPContinuum} presents the 35 high fidelity sources found in the 1.2 mm continuum images. Only one source is barely detected in the natural image and detected with higher signal-to-noise ratio in the tapered image. The source is ASPECS-LP-1mm.C28, a very extended (optical effective radius $r_e>1\arcsec$) spiral galaxy at $z=0.622$. 
Visual inspection revealed that the source ASPECS-LP-1mm.C14 corresponded to what appears to be two galaxies that are situated close to each other. {MUSE and ALMA CO spectroscopy revealed that both galaxies are at similar redshift $z=1.996$ and $z=1.999$, indicating a possible interacting system. Since the two NIR galaxies can clearly be separated (separation of $\sim3\arcsec$ between centroids), ASPECS-LP-1mm.C14 was catalogued in two independent sources, named ASPECS-LP-1mm.C14a and ASPECS-LP-1mm.C14b, for the north-west and south-east galaxy correspondingly.} 

Out of the 35 independent sources, 32 have clear NIR counterparts with measured spectroscopic redshifts. {We define a counterpart as any bright NIR galaxy ($m_{\rm F160W}\leq27$) that is located within the synthesized beam of the natural or tapered image. }
{We stress that out of the 35 source candidates, 26 are catalogued as secure based on the fidelity values. The remaining nine sources are catalogued as candidates based on their fidelity value alone. Despite the latter, the detection of a bright NIR counterpart increases their probability of being a real detection. Based on the measured F160W magnitudes, colors and offsets, we estimate that up to one of the source candidates with NIR counterparts could correspond to a false association.} From the Fidelity values found by LineSeeker, we expect $\sim2.5$ sources to be false. This number is remarkably similar to the number of sources without NIR counterpart, which would suggest that these sources could indeed be false detections. 
{We based this on the fact that our flux density values at 1 mm are at least one order of magnitude lower than the archetypal dark galaxy HDF850.1 \citep{Walter2012} and that dust obscuration is mainly associated with massive galaxies \citep{Whitaker2017}. It is therefore expected that galaxies obscured enough to not be detected in the deep HUDF images should be bright at 1.2mm. Despite that, recent dark galaxies have been discovered with expected 1mm flux density values similar to ASPECS-LP-1mm.C29 \citep{Simpson2014,Williams2019,Wang2019}. We can not discard that the three sources without NIR counterparts are real but the low significance of the detections and their non--detection in deep NIR imaging make them good candidates to be false detections. }
The characterization of the NIR counterpart galaxies of the sources presented here is discussed in a companion paper \citet{Aravena2019b}. We used the Multi-wavelength Analysis of Galaxy Physical Properties \citep[MAGPHYS,][]{daCunha2008,daCunha2015} to estimate galaxy properties such as stellar mass, dust mass and SFR. 

In Tab. \ref{tab:LPContinuum} we also present the flux density values estimated for all the detected sources. We measured the flux densities in the same way as for the ASPECS-LP 3 mm continuum image in \citet{GL2019}. We corrected the flux density measurements for any extended emission detected in the tapered image (see discussion in \S\ref{sec:SourceSearch}) and flux boosting by emission lines (of $\approx30{\rm \ts \mu Jy}$ in the worst case). 
In the case of ASPECS-LP-1mm.C14, we used different apertures to estimate the flux density and corresponding uncertainties associated to each component.

\section{Discussion} \label{sec:Discussion}
\subsection{Number Counts}

\begin{deluxetable*}{cccc}
\tablecaption{Double power-law fit results obtained from the P(D) analysis. \label{tab:DPLFits}}
\tablehead{
\colhead{$N_0$} & 
\colhead{$S_0$} & 
\colhead{$\alpha$} & 
\colhead{$\beta$} \\
\colhead{} & 
\colhead{[mJy]} & 
\colhead{} & 
\colhead{} }
\colnumbers
\startdata
\multicolumn{4}{c}{Differential number counts}\\
$(4.4\pm0.6)\times10^4\ts{\rm mJy^{-1}deg^{-2}}$& $0.10 \pm 0.02$ & $2.5_{-0.1}^{+0.2}$ & $0.0_{-0.2}^{+0.6}$\\
\hline
\multicolumn{4}{c}{Cumulative number counts}\\
$(4.1\pm0.1)\times10^3 \ts{\rm deg^{-2}}$& $0.09\pm0.02$ & $1.94\pm0.14$ & $0.16\pm0.14$
\enddata
\end{deluxetable*}

\begin{deluxetable*}{cccccc}
\tablecaption{ASPECS-LP 1.2mm Differential number counts. \label{tab:1mmDifNumberCounts}}
\tablehead{
\colhead{${\rm S_{\nu}}$ range} & 
\colhead{$\log{\rm S_{\nu}}$} & 
\colhead{${\rm N(S_{\nu})}$} & 
\colhead{${\rm dN/dS}$} & 
\colhead{${\rm\Delta \left(dN/dS\right)}_{Neg}$} & 
\colhead{${\rm\Delta \left(dN/dS\right)}^{Pos}$} \\
\colhead{[$\times 10^{-3}$ mJy]} & 
\colhead{[mJy]} & 
\colhead{} & 
\colhead{$[{\rm mJy^{-1} deg^{-2}}]$} & 
\colhead{$[{\rm mJy^{-1} deg^{-2}}]$} & 
\colhead{$[{\rm mJy^{-1} deg^{-2}}]$}
}
\colnumbers
\startdata
31.6 -- 56.2 & $-$1.38 & 6 & 680000 & 280000 & 340000 \\
56.2 -- 100.0 & $-$1.12 & 9 & 262000 & 88000 & 105000 \\
100.0 -- 177.8 & $-$0.88 & 9 & 124000 & 44000 & 47000 \\
177.8 -- 316.2 & $-$0.62 & 5 & 31000 & 13000 & 16000 \\
316.2 -- 562.3 & $-$0.38 & 4 & 14200 & 6300 & 8400 \\
562.3 -- 1000.0 & $-$0.12 & 1 & 2000 & 1100 & 1900 \\
1000.0 -- 1778.3 & 0.12 & 1 & 1100 & 640 & 1060 \\
\enddata
\tablecomments{
(1) Flux density bin. 
(2) Flux density bin center .
(3) Number of sources per bin (before fidelity and completeness correction). In the case of no sources, an upper limit of $<$1.83 ($1\sigma$ Poisson upper limit for no detection) is used.
(4) Differential number count of sources per square degree. In the case of no sources, a $1\sigma$ upper limit is used. 
(5) Lower uncertainty in the number counts including Poisson errors and flux density errors added in quadrature. 
(6) Upper uncertainty in the number counts including Poisson errors and flux density errors added in quadrature.
}
\end{deluxetable*}

\begin{deluxetable}{cCCccc}
\tablecaption{ASPECS-LP 1.2 mm Cumulative number counts. \label{tab:1mmCumNumberCounts}}
\tablehead{
\colhead{${\rm S_{\nu}}$ range} & 
\colhead{$\log{\rm S_{\nu}}$} & 
\colhead{${\rm N(S_{\nu})}$} & 
\colhead{${\rm N}({\geq \rm S_{\nu}})$} & 
\colhead{$\delta{\rm N}_{Neg}$} & 
\colhead{$\delta{\rm N}^{Pos}$} \\
\colhead{[$\times 10^{-3}$ mJy]} & 
\colhead{[mJy]} & 
\colhead{} & 
\colhead{$[{\rm deg^{-2}}]$} & 
\colhead{$[{\rm deg^{-2}}]$} & 
\colhead{$[{\rm deg^{-2}}]$}
}
\colnumbers
\startdata
31.6 -- 39.8 & $-$1.45 & 4 & 47400 & 8200 & 8900 \\
39.8 -- 50.1 & $-$1.35 & 2 & 35100 & 6200 & 6900 \\
50.1 -- 63.1 & $-$1.25 & 3 & 30700 & 5500 & 6200 \\
63.1 -- 79.4 & $-$1.15 & 0 & 26500 & 4900 & 5600 \\
79.4 -- 100.0 & $-$1.05 & 6 & 26500 & 5100 & 5600 \\
100.0 -- 125.9 & $-$0.95 & 4 & 19300 & 4400 & 4700 \\
125.9 -- 158.5 & $-$0.85 & 4 & 14800 & 3500 & 4000 \\
158.5 -- 199.5 & $-$0.75 & 2 & 10600 & 2900 & 3400 \\
199.5 -- 251.2 & $-$0.65 & 1 & 8700 & 2400 & 3000 \\
251.2 -- 316.2 & $-$0.55 & 3 & 7800 & 2300 & 2700 \\
316.2 -- 398.1 & $-$0.45 & 1 & 5200 & 1900 & 2400 \\
398.1 -- 501.2 & $-$0.35 & 3 & 4300 & 1600 & 2100 \\
501.2 -- 631.0 & $-$0.25 & 0 & 1720 & 840 & 1250 \\
631.0 -- 794.3 & $-$0.15 & 1 & 1720 & 840 & 1250 \\
794.3 -- 1000.0 & $-$0.05 & 0 & 860 & 490 & 820 \\
1000.0 -- 1258.9 & 0.05 & 1 & 860 & 490 & 820 \\
1258.9 -- 1584.9 & 0.15 & 0 & $<$1600 &\nodata&\nodata \\
\enddata
\tablecomments{
(1) Flux density bin. 
(2) Flux density bin center .
(3) Number of sources per bin (before fidelity and completeness correction). In the case of no sources, an upper limit of $<$1.83 ($1\sigma$ Poisson upper limit for no detection) is used.
(4) Cumulative number count of sources per square degree. In the case of no sources, a $1\sigma$ upper limit is used. 
(5) Lower uncertainty in the number counts including Poisson errors and flux density errors added in quadrature. 
(6) Upper uncertainty in the number counts including Poisson errors and flux density errors added in quadrature. 
}
\end{deluxetable}

\begin{figure*}
\epsscale{0.57}
\plotone{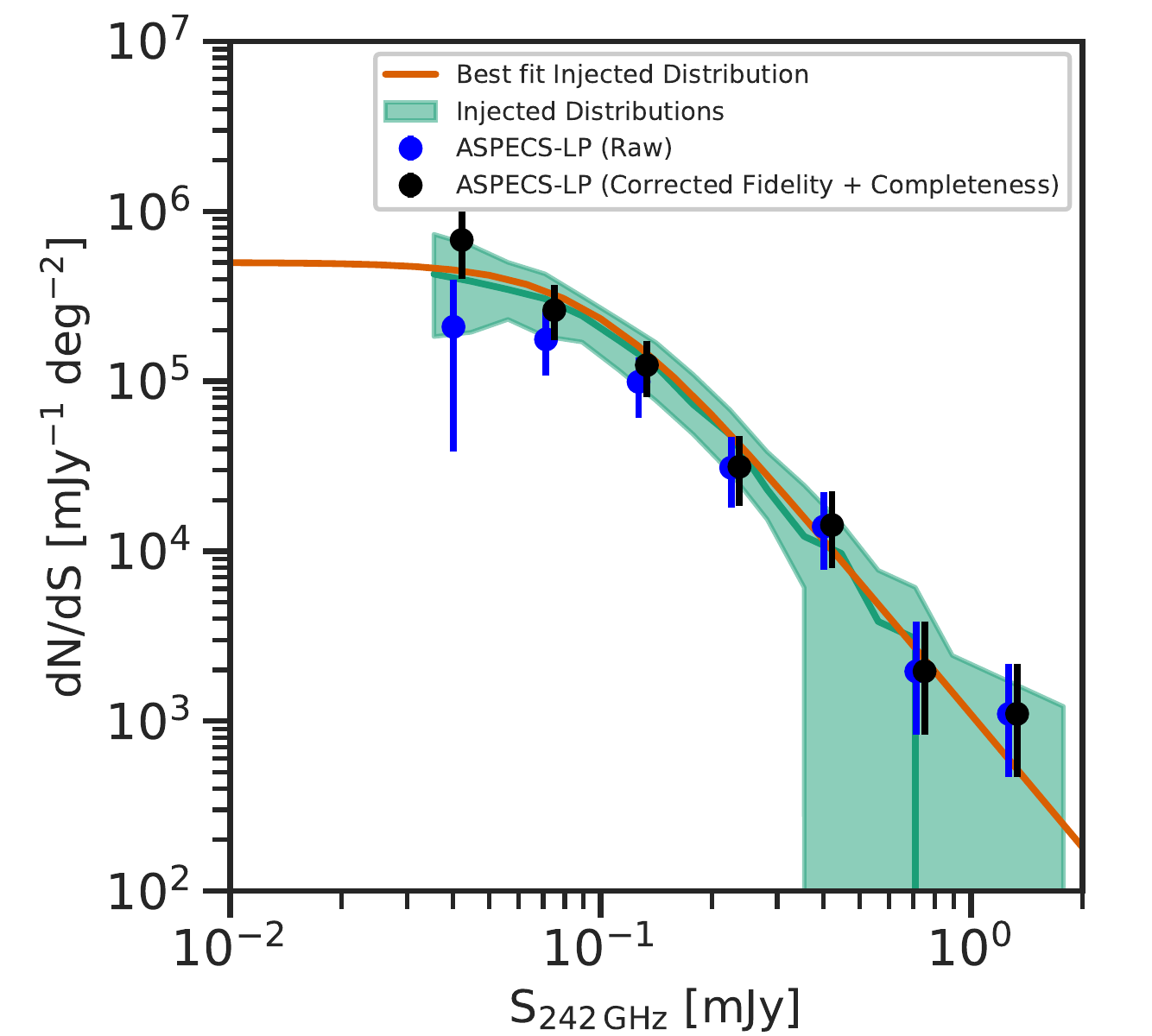}
\plotone{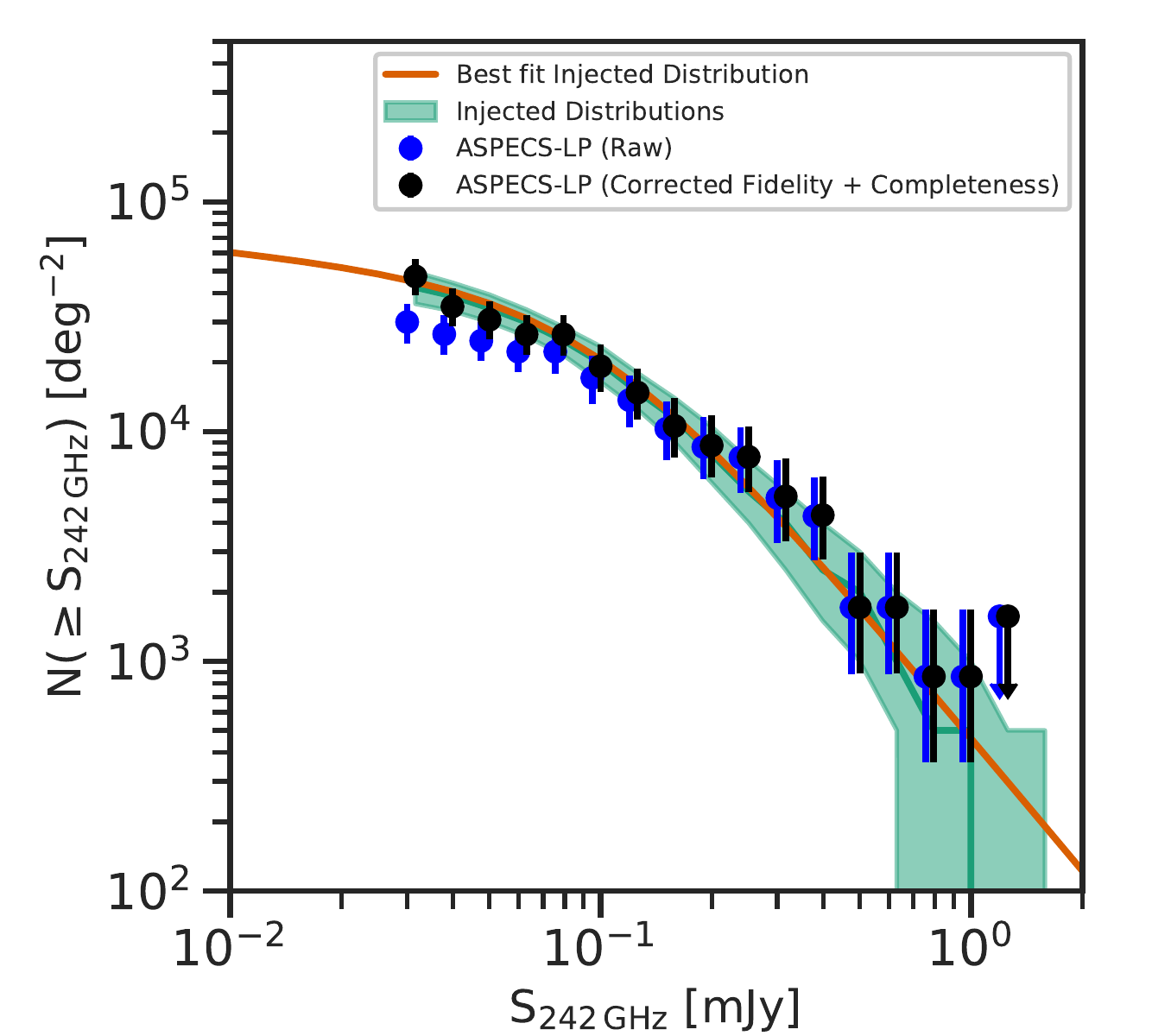}
\plotone{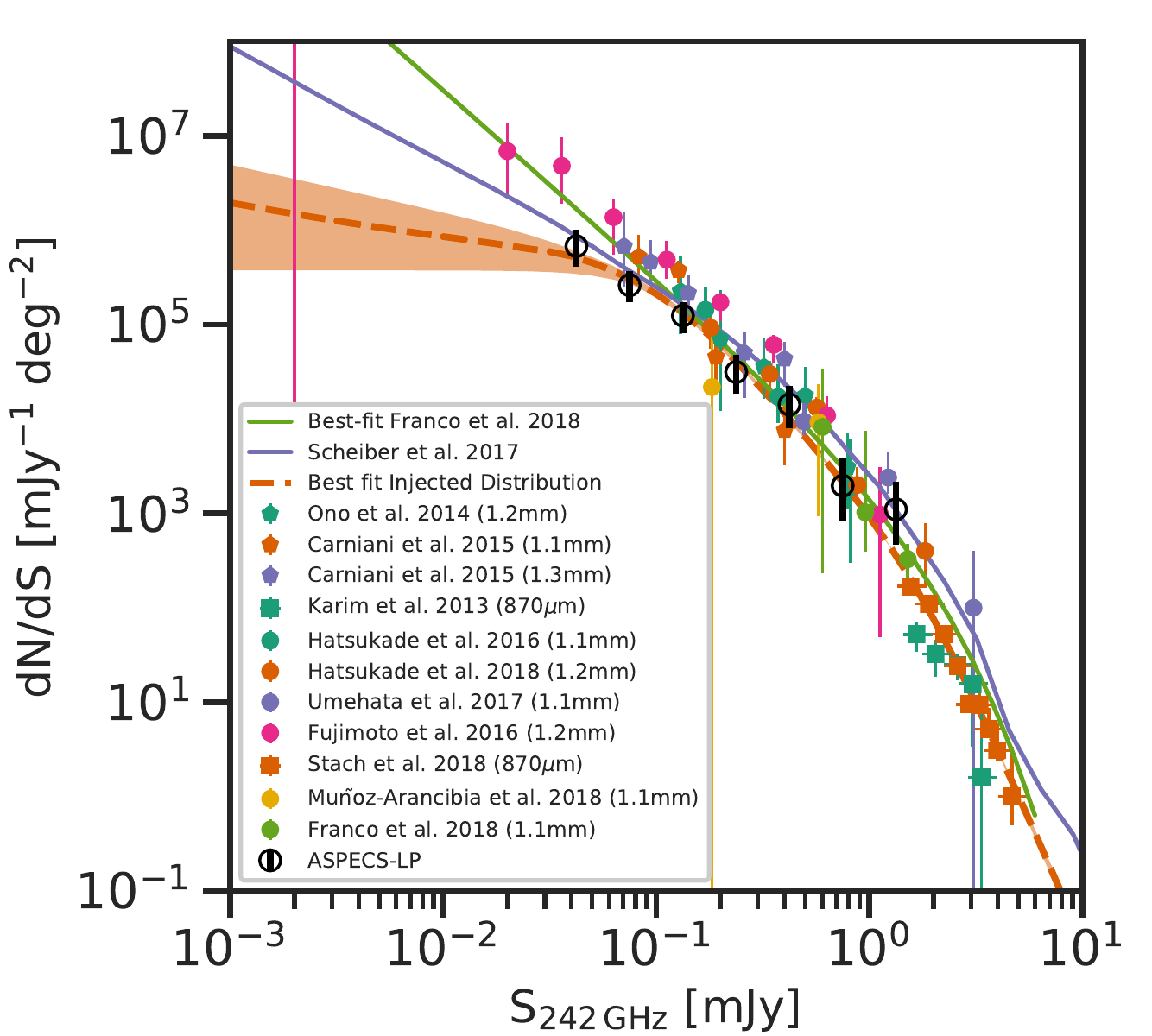}
\plotone{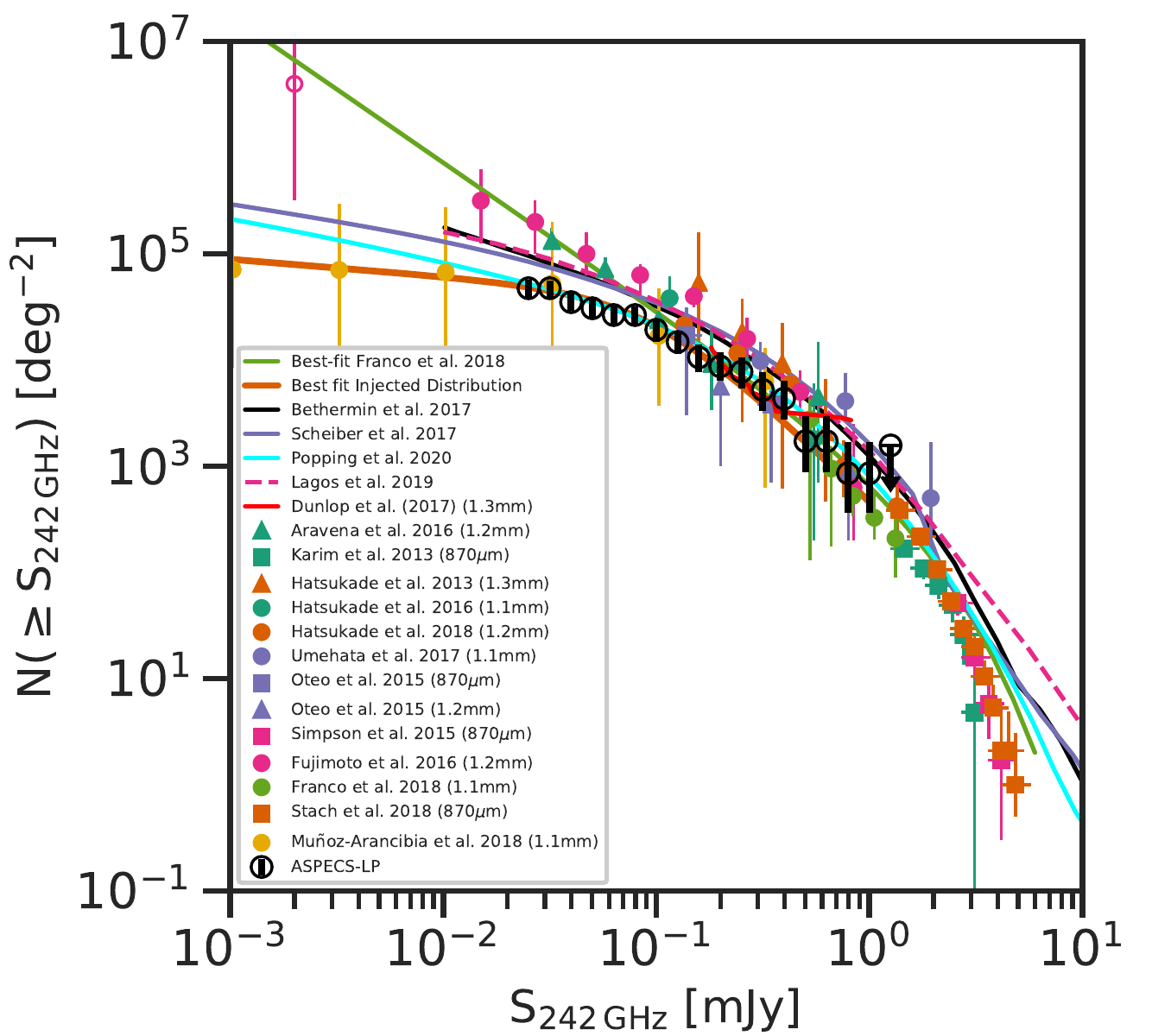}
\caption{The top panels show the number counts (left:differential and right:cumulative) estimates from the direct method (black data points), uncorrected (blue data points) and from the P(D) analysis (green--shaded regions) shown in Fig. \ref{fig:PixelDistribution}. The bottom panels show a comparison to previous studies and models (References in the text). 
The orange lines (dashed and solid) show the best-fit differential counts from the P(D) analysis. In both panels we can see how our results are in good agreement with previous studies at $S_{\nu}>0.1$ mJy but show a clear flattening of the counts (most visible in the cumulative number counts) at $S_{\nu}<0.1$ mJy. The flattening is produced by the change in slope in the differential counts as shown for the orange dashed line. 
\label{fig:NumberCountsResults}}
\end{figure*}

The results from the direct sources number counts as well as the P(D) analysis are presented in Fig. \ref{fig:NumberCountsResults}. The top panels present the uncorrected and corrected number counts (blue and black points respectively) together with the $1\sigma$ range estimated from the forward modeling fitting. The best-fit differential number counts and the double power-law fit to the cumulative number counts resulting from the forward modeling fitting are presented in Tab. \ref{tab:DPLFits}. The best-fit is found by the distribution that minimize $C$ and the $1\sigma$ range is obtained from the probability distributions obtained from the corresponding likelihood values.

Despite the number counts obtained from the P(D) analysis being less well constrained (because of the random positions and corresponding correction by the mosaic PB), the results from the two independent method show excellent agreement. Importantly, the change in slope in the number counts at $S_0\sim0.1\ts{\rm mJy}$ in the differential and cumulative number counts is seen in both analyses. 
The values for the direct number counts are presented in Tab. \ref{tab:1mmDifNumberCounts} and \ref{tab:1mmCumNumberCounts}. 

In the bottom panels of Fig. \ref{fig:NumberCountsResults} we show the direct number counts, estimated as part of this work, and compared to other interferometric studies. The differential number counts are shown in the bottom-left panel. To convert the results from previous studies done at different wavelengths we used a modified black-body and took into account the effects of the cosmic microwave background (CMB) following the recipe presented by \citet{daCunha2013b}. For the dust emissivity index we used a range of $\beta=1.5-2.0$ \citep{DunneAndEales2001,Chapin2009,Clements2010,Draine2011,Planck2011a,Planck2011b}, for the dust temperature we took a range of $25-40$ K \citep{Magdis2012,Magnelli2014,Schreiber2018} and we used a reference redshift of $z=2$. 
The correction factor were defined as $S_{1.2mm} = S_{\lambda}/K_{\lambda}$. The values used are $K_{1.3mm}=0.95$, $K_{1.1mm}=1.3$ and $K_{870\mu m}=2.9\pm0.5$. {These factors correspond to the average of the extreme values obtained for the different properties.} The correction factor for the 870 $\mu m$ has a larger uncertainty given the difference in wavelengths and the intrinsic effects by the range in temperatures. 
The studies at 870 $\mu m$ correspond to follow-up campaigns of single-dish detected sources and from calibration deep fields \citep{Karim2013,Oteo2016,Simpson2015,Stach2018}. These studies mainly sample the bright end of the 1mm galaxy population and correspond to the flux density range of $S_{\nu}>1\ts{\rm mJy}$. 
The studies at 1.1-1.3 millimeters correspond to deep fields and large mosaics in blank and lensing fields \citep{Aravena2016,Hatsukade2013,Hatsukade2016,Umehata2017,Dunlop2017,Hatsukade2018,MunozArancibia2018,Franco2018} as well as the combination of multiple targeted fields \citep{Ono2014,Carniani2015,Oteo2016,Fujimoto2016}. 
Our results are in agreement with the results from other studies at $S_{\nu}>0.1$ mJy but are lower than the rest at fainter flux density values. Our differential number count distribution has a break at $S_{\nu}\sim0.1$ mJy, with the slope towards the faint end being $\beta<0.5$. The orange dashed line shows the best fit obtained as part of the P(D) analysis combined with the extrapolation towards brighter sources. The orange region shows the $1\sigma$ range obtained from the P(D) analysis.

In the bottom-right panel of Fig. \ref{fig:NumberCountsResults} we compare our cumulative number counts with other studies. Compared to most of other studies, and taking into account the scatter between studies, our cumulative number counts in the range 0.1-1 mJy follows the same shape. 
For the counts $S<0.1\ts{\rm mJy}$, our results are considerably below the results from \citet{Fujimoto2016}. We speculate that the difference can be explained by uncertainties in the magnification factors for some of the detections or the usage of targeted fields tracing overdense regions. 
{In the case of the number counts from \citet{Aravena2016}, we reprocessed the same images with the new methods and obtained results consisting with our new results within the error bars. The excess in the counts measured in the ASPECS-Pilot can be attributed to cosmic variance and the different methods used.}

The only other cumulative number counts that seem to agree with our results come from the observations of the ALMA {\it Hubble} Frontier Fields survey \citep{MunozArancibia2018}. In Fig. \ref{fig:NumberCountsResults} we show the updated values using the five galaxy clusters. The Frontier Fields correspond to six galaxy cluster fields observed with multiple observatories with the objective of finding high-redshift galaxies \citep{Lotz2017}. These fields also have the best magnification models ever obtained for galaxy clusters \citep{Priewe2017,Bouwens2017,Meneghetti2017}. \citet{MunozArancibia2018} used a detailed analysis of the source plane reconstruction of the observed images to take into account the effects introduced by the intrinsic sizes of the galaxies and the different lens models used. 
The results from \citet{MunozArancibia2018} have large error bars associated to the intrinsic scatter introduced by the different magnification map models. Despite that, their results fully support our number counts estimates. 

Figure \ref{fig:NumberCountsResults} also presents the 1.2 mm number counts as predicted by different galaxy evolution models \citep{Bethermin2017,Schreiber2017,Lagos2019,Popping2019b}. The four models predict a flattening of the number counts below $S_{\nu}\sim0.1$ mJy with different scaling factors. 

In summary, our results clearly show a flattening of the cumulative number counts and produced by the knee of the differential number counts at $S_{\nu}\sim0.1$ mJy.

\subsection{Number Counts for different populations}

\begin{figure*}
\epsscale{0.57}
\plotone{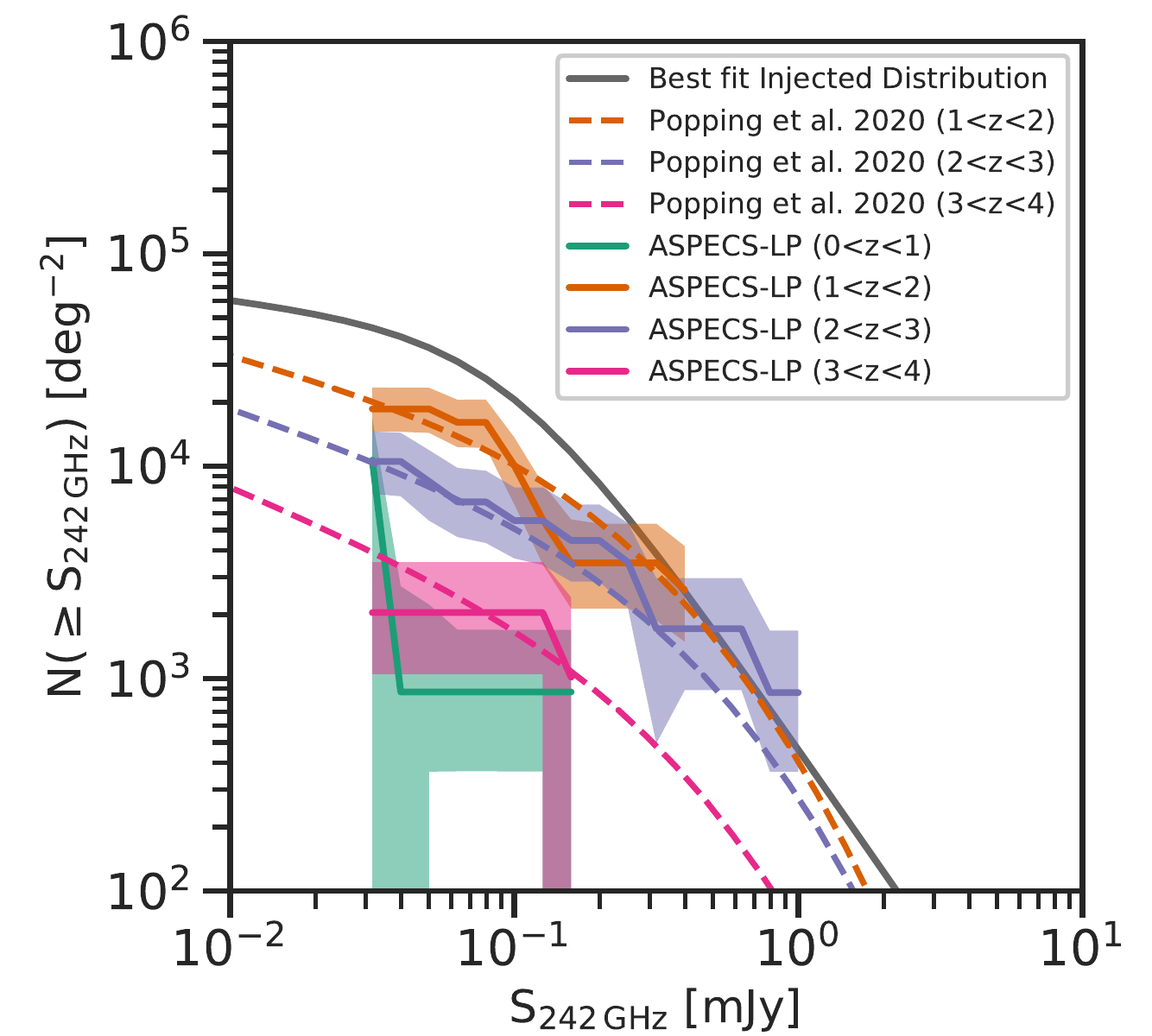}
\plotone{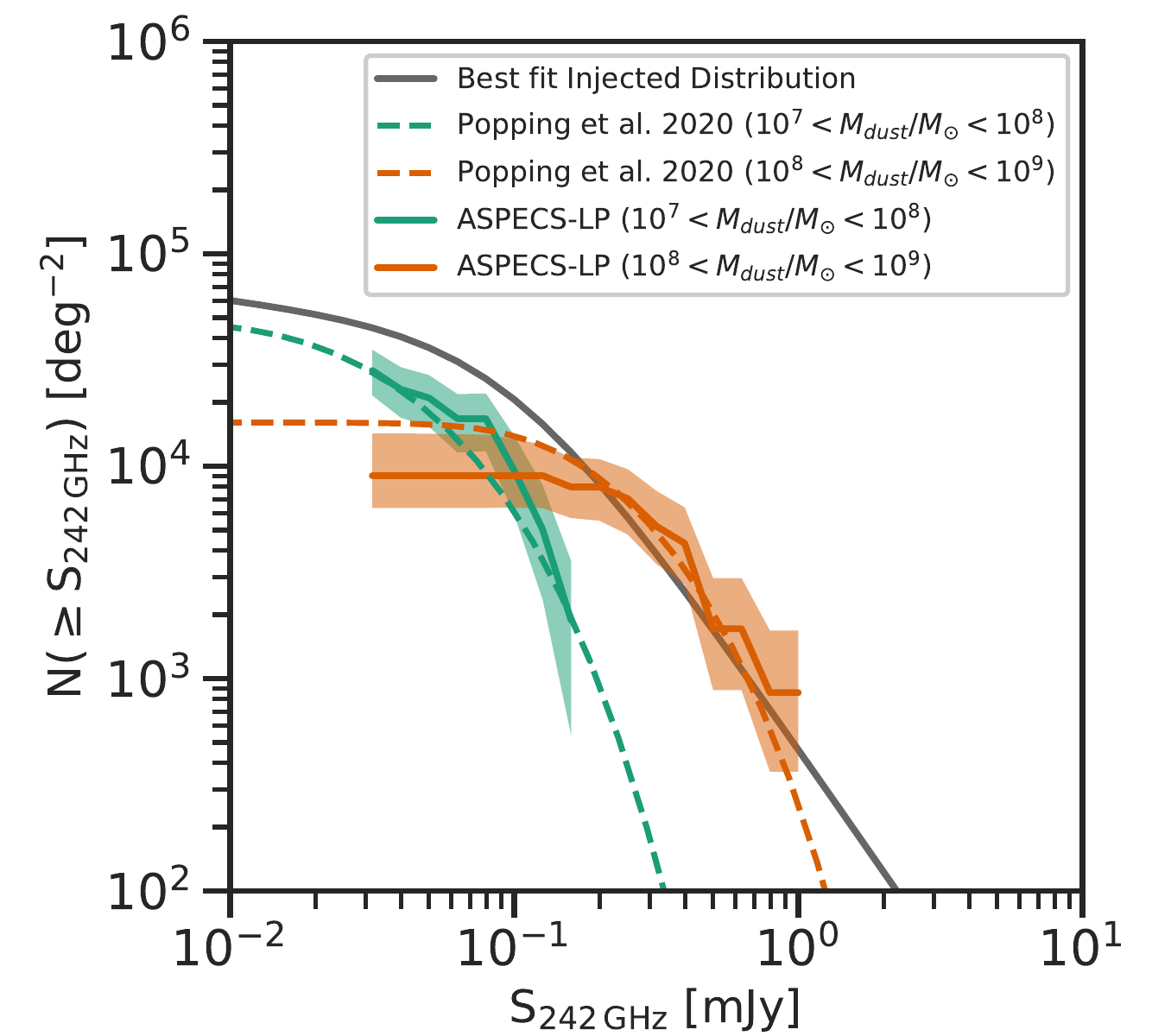}
\plotone{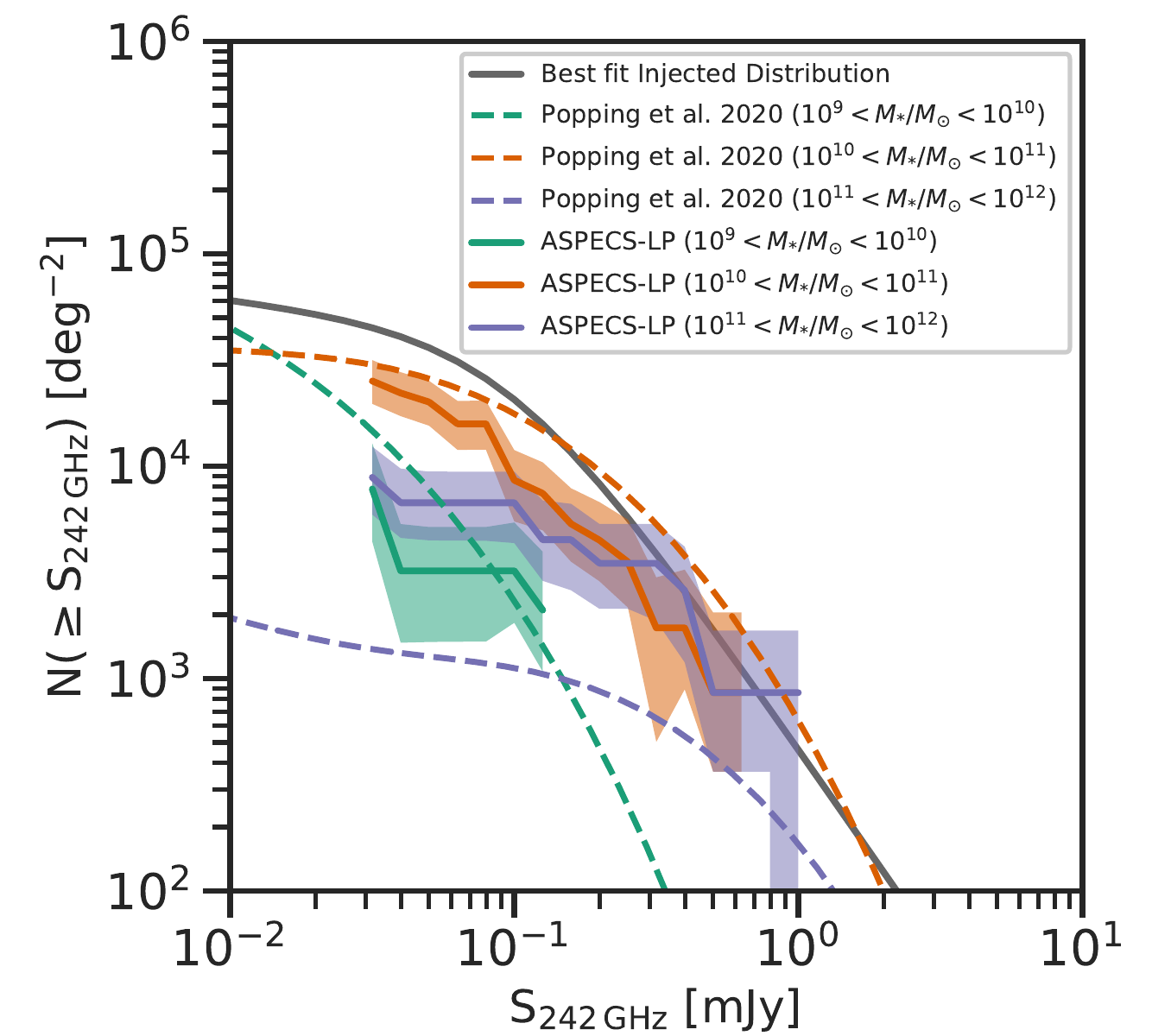}
\plotone{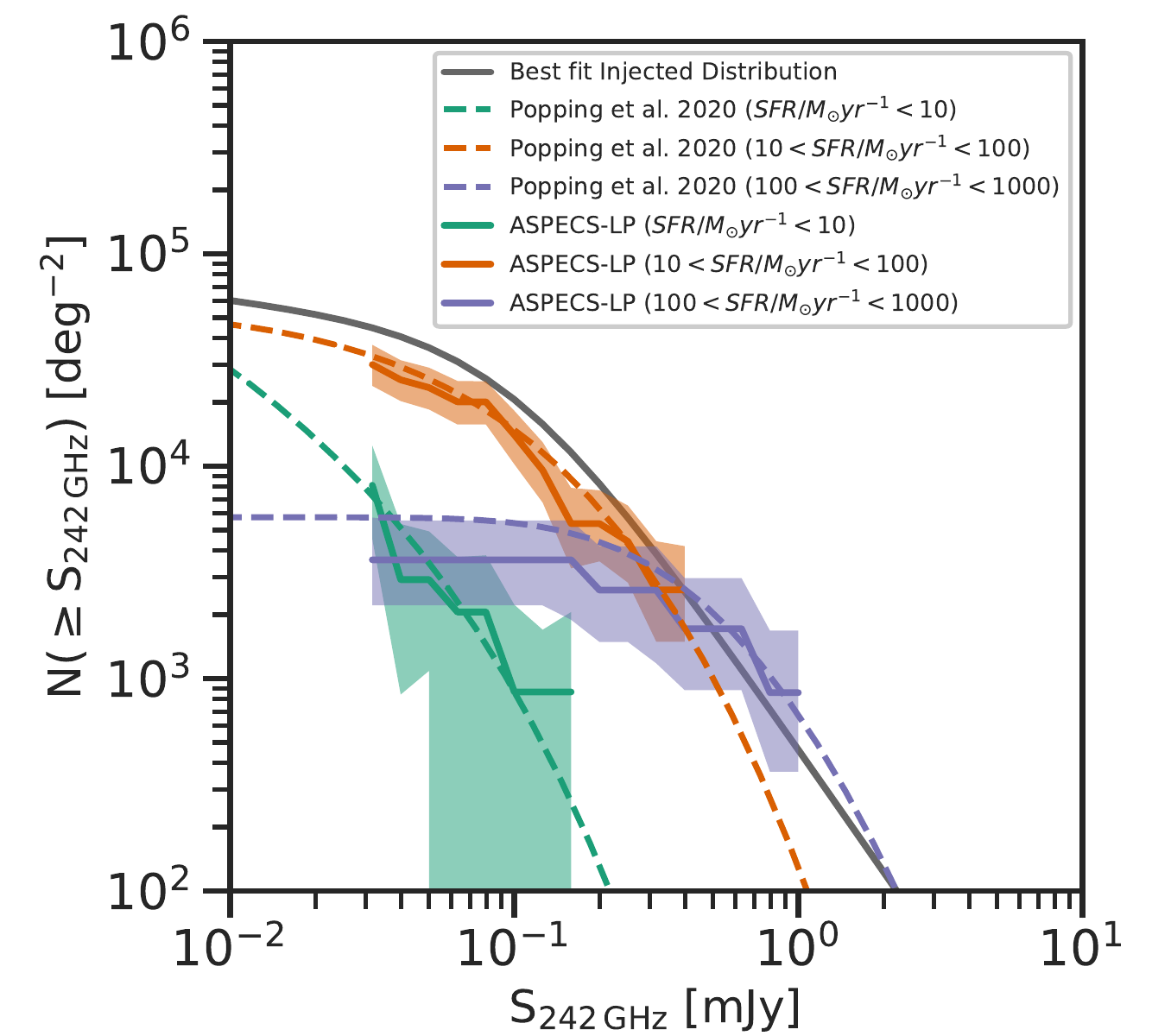}
\caption{Number counts estimated using the detected sources with HST counterparts. We divided the sample based on redshift (top left), dust mass (top right), stellar mass (bottom left) and SFR (bottom right). The solid gray curve is the total cumulative number counts fit shown in Fig. \ref{fig:NumberCountsResults}. The dashed colored lines show the number counts predictions by \citet{Popping2019b}. \label{fig:NumberCountsResultsProperties}}
\end{figure*}

In order to understand the contribution of different galaxy populations to the number counts and thus to the extragalactic background light (EBL) at 1.2 mm, we split the detected sample of dust galaxies into different ranges in stellar mass, SFR, dust mass and redshift using the best SED parameters from the associated optical/NIR counterparts \citep[details in][]{Aravena2019b}.

In Fig. \ref{fig:NumberCountsResultsProperties} we present the number counts for the different populations, split in redshift, dust mass, stellar mass and SFR for the top-left, top-right, bottom-left and bottom-right panels respectively. With this information in hand we can for the first time investigate how different galaxy populations contribute to the number counts. We also show in Fig. \ref{fig:NumberCountsResultsProperties} the corresponding number counts predictions by \citet{Popping2019b}.

{The error bars for the number counts of the different populations are estimated in a similar way as for the whole sample (See \S\ref{sec:DirectNumberCounts}) with the addition of the scatter associated to the estimated properties of the galaxies. We estimate the scatter associated to the uncertainties of the galaxy properties in the same way as for the flux density values. We generate several instances of the galaxy properties based on their best estimates and the corresponding error bars and measure the scatter obtained per bin. The scatter associated to the properties is then added in quadrature to the Poisson statistics error bars and the scatter obtained from the flux density uncertainties. Given a population definition (e.g. range in dust mass) the number counts are estimated assuming this sub-sample as a new sample. We do not include any completeness correction based on the parameters estimated for the galaxies. In other words we assume that the HUDF population is complete down to the levels we are testing. }

When the sample is split by redshift, we note that in most of the flux density range the 1mm selected population is dominated by galaxies in the redshift range of $z=1\text{--}3$. For fainter sources we begin to see the rise of the $z<1$ population. The main component of 1mm galaxy population appears to be between $z=1\text{--}3$, in agreement with the redshift distribution obtained for brighter sources \citep{Simpson2014,Brisbin2017}. It would appear that the population of 1mm galaxies at $z>4$ should be more important at brighter flux density ranges than the ones explored here, at the same time the number density must be such that they are only detectable in large area surveys. The redshift distribution of our sample has a median value that is in agreement with the predictions by \citet{Bethermin2015} based on the depth and wavelength of our observations. See details in \citet{Aravena2019b}.

The sample dust mass has a range between ${\rm M}_{\rm dust}=10^{7}\text{--}10^{9}\msol$. The number counts show that the population with ${\rm M}_{\rm dust}>10^{8}\msol$ is the only component for the sources brighter than 0.2 mJy. For sources fainter than 0.1 mJy the complete population is dominated by the ${\rm M}_{\rm dust}<10^{8}\msol$ population. 

The stellar mass range for our sample is ${\rm M}_{*}=10^{9}\text{--}10^{12}\msol$. All sources brighter than 0.1 mJy have stellar masses ${\rm M}_{*}>10^{10}\msol$.
With the population of 1mm galaxies with stellar mass between ${\rm M}_{*}=10^{10}\text{--}10^{11}\msol$ dominating the counts at all flux density ranges. More massive galaxies, with stellar masses ${\rm M}_{*}>10^{11}\msol$, are responsible for half of the counts in the flux density range 0.3-0.4 mJy, while the 1mm galaxies ${\rm M}_{*}=10^{9}\text{--}10^{10}\msol$ only appear in the faint end of the counts with flux density values below 0.1 mJy.

Finally, when we split the number counts on SFR we see how the brighter 1.2 mm sources are dominated by the population with ${\rm SFR}>100\ts\msol\ts{\rm yr^{-1}}$. In the range below 0.1 mJy we see how the sources with $10\ts\msol\ts{\rm yr^{-1}}<{\rm SFR}<100\ts\msol\ts{\rm yr^{-1}}$ dominate. The faint end of the population also has a small fraction of 1mm galaxies with low SFR rates (${\rm SFR}<10\ts\msol\ts{\rm yr^{-1}}$).

We find that the sources with $S_{\nu}<0.1$ mJy, which correspond to the flattening of the number counts, are galaxies in the redshift range $1<z<2$, dust masses in the range ${\rm M}_{\rm dust}=10^{7}\text{--}10^{8}\msol$, stellar masses in the range ${\rm M}_{*}=10^{10}\text{--}10^{11}\msol$
and SFR in the range $10\ts\msol\ts{\rm yr^{-1}}<{\rm SFR}<100\ts\msol\ts{\rm yr^{-1}}$. 

We remark that the predictions from \citet{Popping2019b} seem to agree fairly well with the observations when split in redshift, dust mass and SFR but not so well when split in stellar mass. The latter disagreement could be explained by an overestimation of the stellar masses for some ASPECS galaxies, since moving sources from the range ${\rm M}_{*}=10^{11}\text{--}10^{12}\msol$ into the lower range would alleviate the disagreement. In fact, we found that the agreement improved when choosing different stellar mass ranges for splitting the number counts, indicating that the disagreement is partially caused by the ranges chosen for the comparison. It has been shown that MAGPHYS can overestimate the stellar mass of galaxies, while providing correct estimates of other properties, when in the presence of AGNs \citep{Hayward2015}. This effect could overestimate the stellar mass for some of our galaxies and produce the disagreement. 

The comparison of the observations with the predictions from \citet{Popping2019b} shows how these new results can be used to test different galaxy evolution models. 
The cumulative number counts for the different populations are presented in Tab. \ref{tab:1mmCumNumberCountsRedshift}, \ref{tab:1mmCumNumberCountsDMass}, \ref{tab:1mmCumNumberCountsStellarMass} and \ref{tab:1mmCumNumberCountsSFR}.  

\subsection{Causes for the flattening of the number counts}

The flattening of the number counts is caused by the absence of a strongly increasing population of galaxies with flux density values lower than $S_{\nu}<0.1$ mJy. We need to identify what types of galaxies could potentially be detected in this flux density range and how they compare with the detected sources.
The companion paper \citet{Popping2019b} discusses in detail the theoretical considerations that predict the flattening of the counts and the properties of the galaxies producing it. According to these models (and in agreement with the results from the previous section) our deep observations are sensitive to galaxies with ${\rm M}_{*}=10^{10}\text{--}10^{11}\msol$ that are on the knee of the stellar and dust mass functions at $1<z<2$. Both mass functions flatten beyond the knee so the number density of galaxies remains almost constant even when going deeper (i.e. observing less massive galaxies). This behaviour is reflected as a flattening of number counts of sources fainter than $S_{\nu}<0.1$ mJy. 
In addition to the number density effect, we are observing galaxies with ${\rm M}_{*}\lesssim10^{10}\msol$ that should have lower gas-phase metallicity than the more massive galaxies and therefore lower dust-to-gas ratios. The dust content, relative to the stellar and gas mass, associated to these galaxies should be lower than the one associated to more massive galaxies. 
The companion paper \citet{Magnelli2019} used stacking to explore the dust content galaxies with different stellar mass ranges. They found that the comoving dust mass density associated to galaxies with ${\rm M}_{*}\lesssim10^{10}\msol$ at $1<z<3$ is fairly low, and that most of the dust mass is associated to more massive galaxies. Similar results are found by \citet{Bouwens2019}, they stacked the emission of all galaxies with ${\rm M}_{*}<10^{9.25}\msol$ at $1.5<z<10$ and found no dust continuum emission. 

In summary, the flattening of the number counts at $S_{\nu}<0.1$ mJy is produced by the lack of continuum emission in galaxies with ${\rm M}_{*}\lesssim10^{10}\msol$ at $1<z<2$. These galaxies are beyond the knee of the stellar and dust mass function, meaning that the number density of galaxies flattens. At the same time, these galaxies are within stellar mass range where the gas-phase metallicity is low enough that the dust content associated to them also decreases, further limiting the number of dust continuum emission detections.

\subsection{1.2 mm extragalactic background light}\label{sec:CIB}

To estimate how much of the 1.2 mm extragalactic background light (EBL) is resolved by our observations we fitted a triple power-law (TPL) to the whole range of observed differential number counts presented in the bottom left panel of Fig. \ref{fig:NumberCountsResults}. We need to use a TPL in order to account for the observed break of the number counts at $\approx0.1$ mJy (Table \ref{tab:DPLFits}) and the break at $\approx1.5$ mJy already identified in wider and shallower observations \citep{Stach2018,Franco2018}.
For the TPL we used the functional form as follow  

\begin{equation}
\label{eq:TPL}
    \frac{dN}{dS} = \left[\frac{1}{F_{1}(S)} + \frac{1}{F_{2}(S)}\right]^{-1},   
\end{equation}
with
\begin{equation}
    F_{1}(S) = \frac{N_{0}}{S_{0}}\left[\left(\frac{S}{S_{0}}\right)^{\alpha}+\left(\frac{S}{S_{0}}\right)^{\beta} \right]^{-1}   
\end{equation}
and 
\begin{equation}
    F_{2}(S) = F_{1}(S_{1})\left(\frac{S}{S_{1}}\right)^{-\gamma}.
\end{equation}

This TPL is a modification of the functional form presented in \citet{Wang2015}. 
For the fitting of TPL we used the ASPECS-LP counts in the range $S_{\nu}<1$ mJy and other studies available for the brighter ones. 
we obtained a best-fit of $S_{1}=1.7\pm0.2\ts{\rm mJy}$ and $\gamma=5.2_{-0.2}^{+0.3}$ when the parameters for $F_1$ follow the $1\sigma$ range from the P(D) analysis. 
We integrated the functional form of the differential number counts (TPL) to obtain the recovered EBL intensity \citep{Franco2018}. We integrated from $35{\rm \ts \mu Jy}$ to infinity and obtained an intensity of $6.3\pm0.2{\rm \ts Jy\ts deg^{-2}}$. We estimated what would be the measurements if we were to detected all the continuum emission in the field. We integrated from zero to infinity and obtained an intensity of $6.8\pm0.4{\rm \ts Jy\ts deg^{-2}}$, which would be our estimate of the total EBL associated to 1mm galaxies in the HUDF.

We compare these values estimated using the best fits of the P(D) with the actual sources extracted in our observations. If we combine the total flux density of the sources presented in Tab. \ref{tab:LPContinuum} we obtain $5.84\pm0.12{\rm \ts Jy\ts deg^{-2}}$ (corrected for fidelity). 
In order to search for possible real continuum source detections skipped by our high fidelity cut we explored the fainter flux density regime. We selected all galaxies with m$_{\rm F160W}<26.4$ in the HUDF and use the corrected-Poissonian probability ($p\leq0.05$) and the 1.2 mm ${\rm S/N}\geq3.0$ positions to select possible real associations \citep{Downes1986,Casey2014}. We complemented the sample by using the deep mid-infrared catalog from \citet{Elbaz2011}. We select all the \textit{Spitzer}-MIPS and \textit{Herschel}-PACS detected sources that show 1.2 mm emission with ${\rm S/N}\geq3.0$. Most of the sources selected by the mid-infrared prior detections are also selected by their HST detection. {The number of source candidates with ${\rm S/N}\geq3.0$ and not selected in the main sample of detection is of $\approx70$ (including sources in the natural and tapered images). The cross-match is done in the same way as for the counterpart search where the allowed offset is given by the corresponding synthesized beam of the detection (See \S\ref{sec:Results}).}
In Tab. \ref{tab:LPContinuumFaint} we present the list of 26 1mm galaxies selected using this method. More details about this complementary sample are presented in \citet{Aravena2019b}.
When we combine the flux density values of those sources we obtain an additional value of $0.93\pm0.07{\rm \ts Jy\ts deg^{-2}}$, combined with the high fidelity sample we obtain thus a total recovered 1.2 mm intensity of $6.77\pm0.14{\rm \ts Jy\ts deg^{-2}}$. This value is in good agreement with the intensity that we should recover in the whole field based on the best fit differential number counts integrating from $35{\rm \ts \mu Jy}$ to infinity. 
In conclusion, by integrating our different number counts and on the secure sources, we recover $\sim93\%$ of the our total estimate for the EBL associated to 1mm galaxies in the HUDF. The recovered emission is closer to $\sim100\%$ when we include the faint sample.

Using our best fit differential number counts we estimated that by going deeper in the same field, to a rms value $\approx5\ts{\rm\mu Jy\ts beam^{-1}}$ (340 hours total time in similar weather conditions as ASPECS-LP), we would recover $\approx99\%$ of our 1.2 mm EBL total estimate. These same observations would only return $\sim5$ extra sources to the already 47 sources discovered (main and secondary samples combined).

Our total estimate for 1.2 mm EBL is lower than previous estimates, our estimated $3\sigma$ upper limit of $<8{\rm \ts Jy\ts deg^{-2}}$ for the HUDF is in tension with the values expected at 242 GHz estimated from Planck observations of $14\text{--}24{\rm \ts Jy\ts deg^{-2}}$ \citep{Planck2014,Odegard2019}. It is important to mention that these estimates are obtained after subtracting the CMB and Galactic dust emission to the observations. This process involved several assumptions and uncertainties, and is not as direct as the count of sources presented here. 
Our $3\sigma$ upper limit is closer, but still lower, than previous estimates of 8--10 ${\rm \ts Jy \ts deg^{-2}}$ \citep{Aravena2016,Oteo2016} and to the values of 10--13 ${\rm \ts Jy\ts deg^{-2}}$ obtained from the modeling of the CIB at multiple wavelengths \citep{Bethermin2012,Khaire2019}. 
In a companion paper, \citet{Popping2019b} estimates the effects of cosmic variance on the number counts measurements. For the wavelength and size of the ASPECS-LP 1.2 mm image they estimate a $2\sigma$ scatter of a factor of 1.5. Such low values of cosmic variance are expected, since the negative K-correction at 1.2 mm allows us to sample a large volume ($z\approx1\text{--}8$) despite the small observed area \citep{Casey2014}. 
Therefore, cosmic variance is not enough to reconcile our estimate of the total 1.2 mm EBL with previous studies.

\subsection{Confusion noise}

In this section we check for the presence of confusion noise in our observations. As stated by several authors \citep{Condon1974,Scheuer1974}, the confusion noise is a source of noise not produced by the instruments or the atmosphere. It is the noise produced by the background of faint unresolved sources that follow a steep distribution \citep{Hogg2001}. The confusion noise is usually estimated as follows

\begin{equation}
\label{eq:Confusion}
    \sigma_c = \sqrt{\sigma_o^2 - \sigma_n^2},  
\end{equation}

with $\sigma_o$ the observed noise in the image and $\sigma_n$ the intrinsic noise of the observations (instrumental + atmospheric) \citep{Condon2012}. In our case, we have an observed value of $\sigma_o=9.3\ts{\rm \mu Jy\ts beam^{-1}}$ (See \S\ref{sec:Observations}) and an estimated $\sigma_n=9.2\ts{\rm \mu Jy\ts beam^{-1}}$ (See \S\ref{sec:P_D_Analysis}). From these values we can estimate a confusion noise of $\sigma_o=1.4\ts{\rm \mu Jy\ts beam^{-1}}$. If we remove the 35 sources discovered as part of this work by fitting a 2D Gaussian emission and measure the rms of the residual image, the value is close to $\sigma_o=9.2\ts{\rm \mu Jy\ts beam^{-1}}$, which would indicate that in fact we are not observing confusion noise but some residuals/sidelobs from real emission that is not properly cleaned. 
The lack of confusion noise in our observations is in line with the flattening of the number counts and the small synthesized beam (see \S\ref{sec:Observations}). 


\section{Conclusion} \label{sec:Conclusion}

In this paper we present the analysis of the deep 1.2 mm continuum image of the HUDF obtained as part of the ASPECS-LP. The image covered an area of 4.2 (2.9) arcmin$^2$ within 10\% (50\%) response of the mosaic primary beam and to a one sigma level of $\sigma=9.3\ts{\rm\mu Jy\ts beam^{-1}}$. With our source extraction methods we recovered 35 significant continuum sources, out of which 32 have clear NIR counterpart galaxies. 
We estimated the number counts by using two independent methods, one using directly the detected sources, corrected by fidelity and completeness. The second method was using the P(D) analysis.

For the P(D) analysis, we use two novel and independent methods to obtain a reliable representation of the observational noise within our image. We made use of Jackknife resampling in the $uv$-plane and in the channel space to obtain continuum noise reference images. The comparison between the different methods showed that the intrinsic noise level of our images is of $\sigma=9.2\ts{\rm\mu Jy\ts beam^{-1}}$.
The P(D) analysis was done by injecting sources to the noise image following different intrinsic differential number counts. The best-fit number counts were found by comparing the pixel distribution of the resulting image with the pixel distribution of the observed dirty image. 
We find that the P(D) analysis number counts are in good agreement with the number counts estimated using the individually detected sources. Importantly, both analysis showed that there is a flattening of the number counts at $S_{\nu}\lesssim0.1$ mJy.

We compare our number counts results with other studies and models. We found a good agreement of our results with other studies at $S_{\nu}>0.1$ mJy but not for the fainter sources. The only other number counts estimates that show a flattening at $S_{\nu}<0.1$ mJy correspond to the ALMA observations of the {\it Hubble} Frontier Fields. 
Our number counts results are in good agreement with the shapes of cumulative number counts predicted using models of galaxy evolution. 
With the detected sources we recovered an intensity of $6.3\pm0.2{\rm \ts Jy\ts deg^{-2}}$, which is $\sim94\%$ of our total estimate of the EBL of $6.8\pm0.4{\rm \ts Jy\ts deg^{-2}}$ for the HUDF. We predicted that doubling the integration time of our observations would only add $\sim5$ extra sources to the ones detected with high fidelity and the ones selected by their counterparts \citep[discussed in][]{Aravena2019b}.

Finally, we presented the number counts for different population of galaxies, split in redshift, dust mass, stellar mass and SFR. These resolved number counts offer a unique opportunity to test our understanding of evolution of galaxies across time. 
We found that the sources with $S_{\nu}<0.1$ mJy (where we detect the flattening of the counts) are dominated by 1mm galaxies in the redshift range $1<z<2$, dust masses in the range ${\rm M}_{\rm dust}=10^{7}\text{--}10^{8}\msol$, stellar masses in the range ${\rm M}_{*}=10^{10}\text{--}10^{11}\msol$
and SFR in the range $10\ts\msol\ts{\rm yr^{-1}}<{\rm SFR}<100\ts\msol\ts{\rm yr^{-1}}$.

\acknowledgments
{We thank the referee for helpful comments that improved this paper.}

This Paper makes use of the ALMA data \newline ADS/JAO.ALMA\#2016.1.00324.L. ALMA is a partnership of ESO (representing its member states), NSF (USA) and NINS (Japan), together with NRC (Canada), NSC and ASIAA (Taiwan), and KASI (Republic of Korea), in cooperation with the Republic of Chile. The Joint ALMA Observatory is operated by ESO, AUI/NRAO and NAOJ. The National Radio Astronomy Observatory is a facility of the National Science Foundation operated under cooperative agreement by Associated Universities, Inc.
``Este trabajo cont\'o con el apoyo de CONICYT + Programa de Astronom\'ia+ Fondo CHINA-CONICYT CAS16026''.
M.N. and F.W. acknowledge support from ERC Advanced Grant 740246 (Cosmic Gas).
D.R. acknowledges support from the National Science Foundation under grant number AST-1614213 and from the Alexander von Humboldt Foundation through a Humboldt Research Fellowship for Experienced Researchers.
RJA was supported by FONDECYT grant number 1191124.
IRS acknowledges support from STFC (ST/P000541/1).
B.M. acknowledge support from the Collaborative Research Centre 956, sub-project A1, funded by the Deutsche Forschungsgemeinschaft (DFG) -- project ID 184018867. 
We acknowledge support from CONICYT grants CATA-Basal AFB-170002 (FEB); 
FONDECYT Regular 1190818 (FEB);
Chile's Ministry of Economy, Development, and Tourism's Millennium Science.
Initiative through grant IC120009, awarded to The Millennium Institute of Astrophysics, MAS (FEB).
Este trabajo cont \'o con el apoyo de CONICYT + PCI + INSTITUTO MAX PLANCK DE ASTRONOMIA MPG190030 (M.A.)

\vspace{5mm}
\facilities{ALMA}

\software{Astropy \citep{Astropy}, \href{https://github.com/jigonzal/LineSeeker}{LineSeeker}
         }

\appendix

\begin{deluxetable*}{ccccccccc}
\tablecaption{Continuum source candidates in ASPECS-LP 1 mm continuum image selected by their HST counterpart. \label{tab:LPContinuumFaint}}
\tablehead{
\colhead{ID} & 
\colhead{R.A.} & 
\colhead{Dec} & 
\colhead{S/N} & 
\colhead{Fidelity} & 
\colhead{PBC} & 
\colhead{$S_{1.2\,\,mm}$} &
\colhead{HST prior} &
\colhead{Mid-IR prior} \\
\colhead{ASPECS-LP.1mm.Faint.} & 
\colhead{} & 
\colhead{} & 
\colhead{} & 
\colhead{} & 
\colhead{} & 
\colhead{[$\rm \mu Jy$]} &
\colhead{} & 
\colhead{} 
}
\colnumbers
\startdata
C01 & 03:32:34.66 & $-$27:47:21.20 & 3.9 & 0.91 & 0.68 & $55.6 \pm 13.7$ & 1& 1\\
C02 & 03:32:35.74 & $-$27:46:39.60 & 3.8 & 0.9 & 0.87 & $41.7 \pm 10.7$ & 1& 1\\
C03 & 03:32:41.32 & $-$27:47:06.60 & 3.7 & 0.91 & 0.95 & $38.0 \pm 9.8$ & 1& 0\\
C04 & 03:32:41.47 & $-$27:47:29.20 & 3.7 & 0.92 & 0.59 & $60.4 \pm 15.8$ & 1& 0\\
C05 & 03:32:37.51 & $-$27:47:56.60 & 3.6 & 0.9 & 0.49 & $70.4 \pm 18.9$ & 1& 0\\
C06 & 03:32:41.63 & $-$27:46:25.80 & 3.6 & 0.9 & 0.45 & $76.2 \pm 20.5$ & 1& 0\\
C07 & 03:32:40.01 & $-$27:47:51.20 & 3.5 & 0.83 & 0.66 & $51.3 \pm 14.0$ & 1& 0\\
C08 & 03:32:35.85 & $-$27:47:18.60 & 3.5 & 0.9 & 0.98 & $34.6 \pm 9.5$ & 1& 1\\
C09 & 03:32:38.56 & $-$27:47:30.60 & 3.4 & 0.9 & 1.0 & $33.3 \pm 9.3$ & 1& 0\\
C10 & 03:32:38.62 & $-$27:47:34.40 & 3.4 & 0.85 & 1.0 & $32.7 \pm 9.3$ & 1& 0\\
C11 & 03:32:36.66 & $-$27:46:31.20 & 3.3 & 0.87 & 0.93 & $34.7 \pm 10.0$ & 1& 1\\
C12 & 03:32:37.17 & $-$27:46:26.20 & 3.3 & 0.85 & 0.95 & $33.4 \pm 9.8$ & 1& 0\\
C13 & 03:32:37.85 & $-$27:47:51.80 & 3.2 & 0.85 & 0.79 & $39.5 \pm 11.7$ & 1& 0\\
C14 & 03:32:35.36 & $-$27:47:17.00 & 3.2 & 0.81 & 0.95 & $32.1 \pm 9.8$ & 1& 0\\
C15 & 03:32:38.36 & $-$27:46:00.20 & 3.1 & 0.81 & 0.69 & $44.1 \pm 13.5$ & 1& 0\\
C16 & 03:32:35.79 & $-$27:46:55.40 & 3.1 & 0.82 & 0.99 & $30.6 \pm 9.4$ & 1& 1\\
C17 & 03:32:38.56 & $-$27:46:31.00 & 3.0 & 0.8 & 0.97 & $34.8 \pm 9.6$ & 1& 0\\
C18 & 03:32:37.32 & $-$27:45:57.80 & 3.0 & 0.8 & 0.47 & $62.5 \pm 19.9$ & 1& 1\\
C19 & 03:32:38.98 & $-$27:46:31.00 & 3.8 & 0.8 & 0.96 & $43.9 \pm 11.7$ & 1& 1\\
C20 & 03:32:39.89 & $-$27:46:07.40 & 3.6 & 0.82 & 0.47 & $85.8 \pm 23.8$ & 1& 1\\
C21 & 03:32:41.35 & $-$27:46:52.00 & 3.5 & 0.84 & 0.95 & $54.0 \pm 15.2$ & 1& 1\\
C22 & 03:32:37.60 & $-$27:47:40.60 & 3.4 & 0.85 & 0.95 & $39.7 \pm 11.9$ & 1& 0\\
C23 & 03:32:42.37 & $-$27:46:57.80 & 3.0 & 0.81 & 0.89 & $38.8 \pm 12.7$ & 1& 1\\
C24 & 03:32:36.86 & $-$27:46:35.00 & 3.0 & 0.82 & 0.96 & $35.9 \pm 11.8$ & 1& 0\\
C25 & 03:32:41.80 & $-$27:47:39.00 & 3.0 & 0.83 & 0.21 & $165.2 \pm 54.2$ & 1& 1\\
C26 & 03:32:38.09 & $-$27:46:14.14 & 3.0 & 0.50 & 0.94 & $39.5 \pm 12.0$ & 0& 1\\
\enddata
\tablecomments{
(1) Identification for continuum source candidates discovered in ASPECS-LP 1.2 mm continuum image.
(2) Right ascension (J2000).
(3) Declination (J2000).
(4) S/N value obtained by LineSeeker assuming an unresolved source.
(5) Fidelity estimate using negative detection and Poisson statistics.
(6) Mosaic primary beam correction.
(7) Integrated flux density at 1.2  mm obtained after removing the channels with bright emission lines when necessary.
(8) If the source is detected based on the HST prior.
(9) If the source is detected based on the mid-infrared prior (\textit{Spitzer}-MIPS and \textit{Herschel}-PACS).
}
\end{deluxetable*}


\startlongtable
\begin{deluxetable*}{cccccc}
\tablecaption{ASPECS-LP 1mm continuum number counts in redshift ranges. \label{tab:1mmCumNumberCountsRedshift}}
\tablehead{
\colhead{${\rm S_{\nu}}$ range} & 
\colhead{$\log{\rm S_{\nu}}$} & 
\colhead{${\rm N(S_{\nu})}$} & 
\colhead{${\rm N}({\geq \rm S_{\nu}})$} & 
\colhead{$\delta{\rm N}_{-}$} & 
\colhead{$\delta{\rm N}^{+}$} \\
\colhead{[$\times 10^{-3}$ mJy]} & 
\colhead{[mJy]} & 
\colhead{} & 
\colhead{$[{\rm deg^{-2}}]$} & 
\colhead{$[{\rm deg^{-2}}]$} & 
\colhead{$[{\rm deg^{-2}}]$}
}
\colnumbers
\startdata
\multicolumn{6}{c}{$0 < z < 1$}\\
\hline
31.6 -- 39.8 & $-$1.45 & 3 & 10700 & 4800 & 6300 \\
39.8 -- 50.1 & $-$1.35 & 0 & 900 & 1700 & 1900 \\
50.1 -- 63.1 & $-$1.25 & 0 & 860 & 500 & 1300 \\
63.1 -- 79.4 & $-$1.15 & 0 & 860 & 500 & 830 \\
79.4 -- 100.0 & $-$1.05 & 0 & 860 & 500 & 830 \\
100.0 -- 125.9 & $-$0.95 & 0 & 860 & 500 & 830 \\
125.9 -- 158.5 & $-$0.85 & 0 & 860 & 500 & 830 \\
158.5 -- 199.5 & $-$0.75 & 1 & 860 & 960 & 830 \\
199.5 -- 251.2 & $-$0.65 & 0 & $<$1600 &\nodata&\nodata \\
\hline
\multicolumn{6}{c}{$1 < z < 2$}\\
\hline
31.6 -- 39.8 & $-$1.45 & 0 & 18600 & 4100 & 4900 \\
39.8 -- 50.1 & $-$1.35 & 0 & 18600 & 4100 & 4800 \\
50.1 -- 63.1 & $-$1.25 & 2 & 18600 & 4200 & 4800 \\
63.1 -- 79.4 & $-$1.15 & 0 & 16100 & 3800 & 4500 \\
79.4 -- 100.0 & $-$1.05 & 5 & 16100 & 3900 & 4500 \\
100.0 -- 125.9 & $-$0.95 & 4 & 10100 & 3400 & 3600 \\
125.9 -- 158.5 & $-$0.85 & 2 & 5600 & 2200 & 2700 \\
158.5 -- 199.5 & $-$0.75 & 0 & 3500 & 1400 & 2100 \\
199.5 -- 251.2 & $-$0.65 & 0 & 3500 & 1400 & 1900 \\
251.2 -- 316.2 & $-$0.55 & 0 & 3500 & 1400 & 1900 \\
316.2 -- 398.1 & $-$0.45 & 1 & 3500 & 1600 & 1900 \\
398.1 -- 501.2 & $-$0.35 & 3 & 2600 & 1100 & 1600 \\
501.2 -- 631.0 & $-$0.25 & 0 & $<$1600 &\nodata&\nodata \\
\hline
\multicolumn{6}{c}{$2 < z < 3$}\\
\hline
31.6 -- 39.8 & $-$1.45 & 0 & 10500 & 3200 & 4000 \\
39.8 -- 50.1 & $-$1.35 & 1 & 10500 & 3300 & 3800 \\
50.1 -- 63.1 & $-$1.25 & 1 & 8500 & 2900 & 3400 \\
63.1 -- 79.4 & $-$1.15 & 0 & 6800 & 2200 & 3000 \\
79.4 -- 100.0 & $-$1.05 & 1 & 6800 & 2400 & 2700 \\
100.0 -- 125.9 & $-$0.95 & 0 & 5500 & 1900 & 2400 \\
125.9 -- 158.5 & $-$0.85 & 1 & 5500 & 2100 & 2400 \\
158.5 -- 199.5 & $-$0.75 & 0 & 4500 & 1600 & 2100 \\
199.5 -- 251.2 & $-$0.65 & 1 & 4500 & 1600 & 2100 \\
251.2 -- 316.2 & $-$0.55 & 2 & 3500 & 1400 & 1900 \\
316.2 -- 398.1 & $-$0.45 & 0 & 1700 & 1200 & 1300 \\
398.1 -- 501.2 & $-$0.35 & 0 & 1720 & 840 & 1250 \\
501.2 -- 631.0 & $-$0.25 & 0 & 1720 & 840 & 1250 \\
631.0 -- 794.3 & $-$0.15 & 1 & 1720 & 840 & 1250 \\
794.3 -- 1000.0 & $-$0.05 & 0 & 860 & 490 & 820 \\
1000.0 -- 1258.9 & 0.05 & 1 & 860 & 490 & 820 \\
1258.9 -- 1584.9 & 0.15 & 0 & $<$1600 &\nodata&\nodata \\
\hline
\multicolumn{6}{c}{$3 < z < 4$}\\
\hline
31.6 -- 39.8 & $-$1.45 & 0 & 2040 & 1000 & 1490 \\
39.8 -- 50.1 & $-$1.35 & 0 & 2040 & 1000 & 1490 \\
50.1 -- 63.1 & $-$1.25 & 0 & 2040 & 1000 & 1490 \\
63.1 -- 79.4 & $-$1.15 & 0 & 2040 & 1000 & 1490 \\
79.4 -- 100.0 & $-$1.05 & 0 & 2040 & 1000 & 1490 \\
100.0 -- 125.9 & $-$0.95 & 0 & 2040 & 1000 & 1490 \\
125.9 -- 158.5 & $-$0.85 & 1 & 2040 & 1000 & 1490 \\
158.5 -- 199.5 & $-$0.75 & 1 & 1000 & 1200 & 1400 \\
199.5 -- 251.2 & $-$0.65 & 0 & $<$1600 &\nodata&\nodata \\
\hline
\enddata
\tablecomments{
(1) Flux density bin. 
(2) Flux density bin center .
(3) Number of sources per bin (before fidelity and completeness correction). In the case of no sources, an upper limit of $<$1.83 is used.
(4) Cumulative number count of sources per square degree. In the case of no sources, a $1\sigma$ upper limit is used. 
(5) Lower uncertainty in the number counts.
(6) Upper uncertainty in the number counts.
}
\end{deluxetable*}

\startlongtable
\begin{deluxetable*}{cccccc}
\tablecaption{ASPECS-LP 1mm continuum number counts in dust mass ranges. \label{tab:1mmCumNumberCountsDMass}}
\tablehead{
\colhead{${\rm S_{\nu}}$ range} & 
\colhead{$\log{\rm S_{\nu}}$} & 
\colhead{${\rm N(S_{\nu})}$} & 
\colhead{${\rm N}({\geq \rm S_{\nu}})$} & 
\colhead{$\delta{\rm N}_{-}$} & 
\colhead{$\delta{\rm N}^{+}$} \\
\colhead{[$\times 10^{-3}$ mJy]} & 
\colhead{[mJy]} & 
\colhead{} & 
\colhead{$[{\rm deg^{-2}}]$} & 
\colhead{$[{\rm deg^{-2}}]$} & 
\colhead{$[{\rm deg^{-2}}]$}
}
\colnumbers
\startdata
\multicolumn{6}{c}{$10^{7}<M_{dust}/M_{\odot}<10^{8}$}\\
\hline
31.6 -- 39.8 & $-$1.45 & 2 & 28200 & 6700 & 7200 \\
39.8 -- 50.1 & $-$1.35 & 1 & 23000 & 5800 & 6200 \\
50.1 -- 63.1 & $-$1.25 & 3 & 21000 & 5700 & 5800 \\
63.1 -- 79.4 & $-$1.15 & 0 & 16700 & 5300 & 5100 \\
79.4 -- 100.0 & $-$1.05 & 6 & 16700 & 5100 & 5100 \\
100.0 -- 125.9 & $-$0.95 & 4 & 9500 & 4000 & 4000 \\
125.9 -- 158.5 & $-$0.85 & 3 & 5000 & 2700 & 3100 \\
158.5 -- 199.5 & $-$0.75 & 2 & 1900 & 1300 & 1700 \\
199.5 -- 251.2 & $-$0.65 & 0 & $<$1600 &\nodata&\nodata \\
\hline
\multicolumn{6}{c}{$10^{8}<M_{dust}/M_{\odot}<10^{9}$}\\
\hline
31.6 -- 39.8 & $-$1.45 & 0 & 9000 & 2700 & 5400 \\
39.8 -- 50.1 & $-$1.35 & 0 & 9000 & 2700 & 5300 \\
50.1 -- 63.1 & $-$1.25 & 0 & 9000 & 2700 & 5300 \\
63.1 -- 79.4 & $-$1.15 & 0 & 9000 & 2700 & 5300 \\
79.4 -- 100.0 & $-$1.05 & 0 & 9000 & 2700 & 5200 \\
100.0 -- 125.9 & $-$0.95 & 0 & 9000 & 2700 & 4400 \\
125.9 -- 158.5 & $-$0.85 & 1 & 9000 & 2700 & 3600 \\
158.5 -- 199.5 & $-$0.75 & 0 & 8000 & 2300 & 3000 \\
199.5 -- 251.2 & $-$0.65 & 1 & 8000 & 2400 & 2800 \\
251.2 -- 316.2 & $-$0.55 & 2 & 7000 & 2300 & 2600 \\
316.2 -- 398.1 & $-$0.45 & 1 & 5200 & 1800 & 2400 \\
398.1 -- 501.2 & $-$0.35 & 3 & 4300 & 1600 & 2100 \\
501.2 -- 631.0 & $-$0.25 & 0 & 1720 & 840 & 1250 \\
631.0 -- 794.3 & $-$0.15 & 1 & 1720 & 840 & 1250 \\
794.3 -- 1000.0 & $-$0.05 & 0 & 860 & 490 & 820 \\
1000.0 -- 1258.9 & 0.05 & 1 & 860 & 490 & 820 \\
1258.9 -- 1584.9 & 0.15 & 0 & $<$1600 &\nodata&\nodata \\
\hline
\enddata
\tablecomments{
(1) Flux density bin. 
(2) Flux density bin center .
(3) Number of sources per bin (before fidelity and completeness correction). In the case of no sources, an upper limit of $<$1.83 is used.
(4) Cumulative number count of sources per square degree. In the case of no sources, a $1\sigma$ upper limit is used. 
(5) Lower uncertainty in the number counts.
(6) Upper uncertainty in the number counts.
}
\end{deluxetable*}

\startlongtable
\begin{deluxetable*}{cccccc}
\tablecaption{ASPECS-LP 1mm continuum number counts in Stellar mass ranges. \label{tab:1mmCumNumberCountsStellarMass}}
\tablehead{
\colhead{${\rm S_{\nu}}$ range} & 
\colhead{$\log{\rm S_{\nu}}$} & 
\colhead{${\rm N(S_{\nu})}$} & 
\colhead{${\rm N}({\geq \rm S_{\nu}})$} & 
\colhead{$\delta{\rm N}_{-}$} & 
\colhead{$\delta{\rm N}^{+}$} \\
\colhead{[$\times 10^{-3}$ mJy]} & 
\colhead{[mJy]} & 
\colhead{} & 
\colhead{$[{\rm deg^{-2}}]$} & 
\colhead{$[{\rm deg^{-2}}]$} & 
\colhead{$[{\rm deg^{-2}}]$}
}
\colnumbers
\startdata
\multicolumn{6}{c}{$10^9<M_{*}/M_{\odot}<10^{10}$}\\
\hline
31.6 -- 39.8 & $-$1.45 & 1 & 7800 & 3400 & 4900 \\
39.8 -- 50.1 & $-$1.35 & 0 & 3200 & 1700 & 2300 \\
50.1 -- 63.1 & $-$1.25 & 0 & 3200 & 1700 & 2000 \\
63.1 -- 79.4 & $-$1.15 & 0 & 3200 & 1700 & 2000 \\
79.4 -- 100.0 & $-$1.05 & 0 & 3200 & 1700 & 2000 \\
100.0 -- 125.9 & $-$0.95 & 1 & 3200 & 1400 & 2200 \\
125.9 -- 158.5 & $-$0.85 & 2 & 2100 & 1000 & 1900 \\
158.5 -- 199.5 & $-$0.75 & 0 & $<$1600 &\nodata&\nodata \\
\hline
\multicolumn{6}{c}{$10^{10}<M_{*}/M_{\odot}<10^{11}$}\\
\hline
31.6 -- 39.8 & $-$1.45 & 1 & 25200 & 5600 & 6300 \\
39.8 -- 50.1 & $-$1.35 & 1 & 22100 & 4900 & 5600 \\
50.1 -- 63.1 & $-$1.25 & 3 & 20000 & 4500 & 5200 \\
63.1 -- 79.4 & $-$1.15 & 0 & 15800 & 3900 & 4600 \\
79.4 -- 100.0 & $-$1.05 & 6 & 15800 & 3900 & 4500 \\
100.0 -- 125.9 & $-$0.95 & 1 & 8600 & 3100 & 3300 \\
125.9 -- 158.5 & $-$0.85 & 2 & 7400 & 2500 & 3000 \\
158.5 -- 199.5 & $-$0.75 & 1 & 5300 & 2000 & 2500 \\
199.5 -- 251.2 & $-$0.65 & 1 & 4500 & 1600 & 2300 \\
251.2 -- 316.2 & $-$0.55 & 2 & 3500 & 1400 & 2100 \\
316.2 -- 398.1 & $-$0.45 & 0 & 1700 & 1200 & 1300 \\
398.1 -- 501.2 & $-$0.35 & 1 & 1730 & 840 & 1520 \\
501.2 -- 631.0 & $-$0.25 & 0 & 860 & 490 & 1190 \\
631.0 -- 794.3 & $-$0.15 & 1 & 860 & 490 & 1190 \\
794.3 -- 1000.0 & $-$0.05 & 0 & $<$1600 &\nodata&\nodata \\
\hline
\multicolumn{6}{c}{$10^{11}<M_{*}/M_{\odot}<10^{12}$}\\
\hline
31.6 -- 39.8 & $-$1.45 & 1 & 8900 & 2900 & 3500 \\
39.8 -- 50.1 & $-$1.35 & 0 & 6700 & 2100 & 3000 \\
50.1 -- 63.1 & $-$1.25 & 0 & 6700 & 2300 & 2700 \\
63.1 -- 79.4 & $-$1.15 & 0 & 6700 & 2300 & 2700 \\
79.4 -- 100.0 & $-$1.05 & 0 & 6700 & 2300 & 2700 \\
100.0 -- 125.9 & $-$0.95 & 2 & 6700 & 2400 & 2700 \\
125.9 -- 158.5 & $-$0.85 & 0 & 4500 & 1600 & 2400 \\
158.5 -- 199.5 & $-$0.75 & 1 & 4500 & 1900 & 2100 \\
199.5 -- 251.2 & $-$0.65 & 0 & 3500 & 1400 & 1900 \\
251.2 -- 316.2 & $-$0.55 & 0 & 3500 & 1400 & 1900 \\
316.2 -- 398.1 & $-$0.45 & 1 & 3500 & 1600 & 1900 \\
398.1 -- 501.2 & $-$0.35 & 2 & 2600 & 1400 & 1600 \\
501.2 -- 631.0 & $-$0.25 & 0 & 860 & 490 & 820 \\
631.0 -- 794.3 & $-$0.15 & 0 & 860 & 490 & 820 \\
794.3 -- 1000.0 & $-$0.05 & 0 & 860 & 490 & 820 \\
1000.0 -- 1258.9 & 0.05 & 1 & 860 & 990 & 820 \\
1258.9 -- 1584.9 & 0.15 & 0 & $<$1600 &\nodata&\nodata \\
\hline
\enddata
\tablecomments{
(1) Flux density bin. 
(2) Flux density bin center .
(3) Number of sources per bin (before fidelity and completeness correction). In the case of no sources, an upper limit of $<$1.83 is used.
(4) Cumulative number count of sources per square degree. In the case of no sources, a $1\sigma$ upper limit is used. 
(5) Lower uncertainty in the number counts.
(6) Upper uncertainty in the number counts.
}
\end{deluxetable*}

\startlongtable
\begin{deluxetable*}{cccccc}
\tablecaption{ASPECS-LP 1mm continuum number counts in star-formation rate ranges. \label{tab:1mmCumNumberCountsSFR}}
\tablehead{
\colhead{${\rm S_{\nu}}$ range} & 
\colhead{$\log{\rm S_{\nu}}$} & 
\colhead{${\rm N(S_{\nu})}$} & 
\colhead{${\rm N}({\geq \rm S_{\nu}})$} & 
\colhead{$\delta{\rm N}_{-}$} & 
\colhead{$\delta{\rm N}^{+}$} \\
\colhead{[$\times 10^{-3}$ mJy]} & 
\colhead{[mJy]} & 
\colhead{} & 
\colhead{$[{\rm deg^{-2}}]$} & 
\colhead{$[{\rm deg^{-2}}]$} & 
\colhead{$[{\rm deg^{-2}}]$}
}
\colnumbers
\startdata
\multicolumn{6}{c}{$SFR/M_{\odot}yr^{-1}<10$}\\
\hline
31.6 -- 39.8 & $-$1.45 & 2 & 8100 & 3700 & 4500 \\
39.8 -- 50.1 & $-$1.35 & 0 & 2900 & 2000 & 2500 \\
50.1 -- 63.1 & $-$1.25 & 1 & 2900 & 1800 & 2000 \\
63.1 -- 79.4 & $-$1.15 & 0 & 2100 & 2200 & 1700 \\
79.4 -- 100.0 & $-$1.05 & 1 & 2100 & 1600 & 1700 \\
100.0 -- 125.9 & $-$0.95 & 0 & 860 & 990 & 1380 \\
125.9 -- 158.5 & $-$0.85 & 0 & 860 & 960 & 830 \\
158.5 -- 199.5 & $-$0.75 & 1 & 860 & 500 & 1190 \\
199.5 -- 251.2 & $-$0.65 & 0 & $<$1600 &\nodata&\nodata \\
\hline
\multicolumn{6}{c}{$10<SFR/M_{\odot}yr^{-1}<100$}\\
\hline
31.6 -- 39.8 & $-$1.45 & 1 & 30100 & 6200 & 7100 \\
39.8 -- 50.1 & $-$1.35 & 1 & 25500 & 5300 & 6000 \\
50.1 -- 63.1 & $-$1.25 & 2 & 23400 & 4900 & 5600 \\
63.1 -- 79.4 & $-$1.15 & 0 & 20100 & 4400 & 5000 \\
79.4 -- 100.0 & $-$1.05 & 5 & 20100 & 4400 & 5000 \\
100.0 -- 125.9 & $-$0.95 & 4 & 14000 & 3800 & 4200 \\
125.9 -- 158.5 & $-$0.85 & 4 & 9600 & 2800 & 3400 \\
158.5 -- 199.5 & $-$0.75 & 0 & 5400 & 2100 & 2500 \\
199.5 -- 251.2 & $-$0.65 & 1 & 5400 & 1800 & 2300 \\
251.2 -- 316.2 & $-$0.55 & 2 & 4400 & 1800 & 2100 \\
316.2 -- 398.1 & $-$0.45 & 0 & 2600 & 1400 & 1600 \\
398.1 -- 501.2 & $-$0.35 & 3 & 2600 & 1100 & 1600 \\
501.2 -- 631.0 & $-$0.25 & 0 & $<$1600 &\nodata&\nodata \\
\hline
\multicolumn{6}{c}{$100<SFR/M_{\odot}yr^{-1}1000$}\\
\hline
31.6 -- 39.8 & $-$1.45 & 0 & 3600 & 1400 & 2100 \\
39.8 -- 50.1 & $-$1.35 & 0 & 3600 & 1400 & 2100 \\
50.1 -- 63.1 & $-$1.25 & 0 & 3600 & 1400 & 2100 \\
63.1 -- 79.4 & $-$1.15 & 0 & 3600 & 1400 & 2100 \\
79.4 -- 100.0 & $-$1.05 & 0 & 3600 & 1400 & 2100 \\
100.0 -- 125.9 & $-$0.95 & 0 & 3600 & 1400 & 2100 \\
125.9 -- 158.5 & $-$0.85 & 0 & 3600 & 1400 & 1900 \\
158.5 -- 199.5 & $-$0.75 & 1 & 3600 & 1700 & 1900 \\
199.5 -- 251.2 & $-$0.65 & 0 & 2600 & 1100 & 1600 \\
251.2 -- 316.2 & $-$0.55 & 0 & 2600 & 1100 & 1600 \\
316.2 -- 398.1 & $-$0.45 & 1 & 2600 & 1400 & 1600 \\
398.1 -- 501.2 & $-$0.35 & 0 & 1720 & 840 & 1250 \\
501.2 -- 631.0 & $-$0.25 & 0 & 1720 & 840 & 1250 \\
631.0 -- 794.3 & $-$0.15 & 1 & 1720 & 840 & 1250 \\
794.3 -- 1000.0 & $-$0.05 & 0 & 860 & 490 & 820 \\
1000.0 -- 1258.9 & 0.05 & 1 & 860 & 490 & 820 \\
1258.9 -- 1584.9 & 0.15 & 0 & $<$1600 &\nodata&\nodata \\
\enddata
\tablecomments{
(1) Flux density bin. 
(2) Flux density bin center .
(3) Number of sources per bin (before fidelity and completeness correction). In the case of no sources, an upper limit of $<$1.83 is used.
(4) Cumulative number count of sources per square degree. In the case of no sources, a $1\sigma$ upper limit is used. 
(5) Lower uncertainty in the number counts.
(6) Upper uncertainty in the number counts.
}
\end{deluxetable*}

\end{document}